\documentclass[iop,numberedappendix]{emulateapj}

\usepackage{verbatim}
\usepackage{amsmath}

\def\M{{\cal M}}

\newcommand{\simgt}{\,\hbox{\lower0.6ex\hbox{$\sim$}\llap{\raise0.6ex\hbox{$>$}}}\,}
\newcommand{\simlt}{\,\hbox{\lower0.6ex\hbox{$\sim$}\llap{\raise0.6ex\hbox{$<$}}}\,}

\begin{document}

\title{Sloshing of the Magnetized Cool Gas in the Cores of Galaxy Clusters}

\author{J. A. ZuHone\altaffilmark{1,2}, M. Markevitch\altaffilmark{1,2}, D. Lee\altaffilmark{3}}

\altaffiltext{1}{Astrophysics Science Division, Laboratory for High Energy Astrophysics, Code 662, NASA/Goddard Space Flight Center, Greenbelt, MD 20771}
\altaffiltext{2}{Smithsonian Astrophysical Observatory, Harvard-Smithsonian Center for Astrophysics, Cambridge, MA 02138}
\altaffiltext{3}{Department of Astronomy, ASC Flash Center, University of Chicago, 5747 S. Ellis Avenue, Chicago, IL 60637}

\begin{abstract}
X-ray observations of many clusters of galaxies reveal the presence of edges in surface brightness and temperature, known as ``cold fronts''. In relaxed clusters with cool cores, these edges have been interpreted as evidence for the ``sloshing'' of the core gas in the cluster's gravitational potential. The smoothness of these edges has been interpreted as evidence for the stabilizing effect of magnetic fields ``draped'' around the front surfaces. To check this hypothesis, we perform high-resolution magnetohydrodynamics simulations of magnetized gas sloshing in galaxy clusters initiated by encounters with subclusters. We go beyond previous works on the simulation of cold fronts in a magnetized intracluster medium by simulating their formation in realistic, idealized mergers with high resolution ($\Delta{x} \sim 2$~kpc). Our simulations sample a parameter space of plausible initial magnetic field strengths and field configurations. In the simulations, we observe strong velocity shears associated with the cold fronts amplifying the magnetic field along the cold front surfaces, increasing the magnetic field strength in these layers by up to an order of magnitude, and boosting the magnetic pressure up to near-equipartition with thermal pressure in some cases. In these layers, the magnetic field becomes strong enough to stabilize the cold fronts against Kelvin-Helmholtz instabilities, resulting in sharp, smooth fronts as those seen in observations of real clusters. These magnetic fields also result in strong suppression of mixing of high and low-entropy gas in the cluster, seen in our simulations of mergers in the absence of a magnetic field. As a result, the heating of the core due to sloshing is very modest and is unable to stave off a cooling catastrophe. 
\end{abstract}


\section{Introduction\label{sec:intro}}

Observations with the most recent generation of X-ray telescopes have shown that clusters looking ``relaxed'' upon first glance often have cool cores that are in a disturbed state. Many ``cool-core'' systems exhibit edges in X-ray surface brightness approximately concentric with respect to brightness peak of the cluster \citep[e.g.,][]{maz01,mar01,MVF03}. X-ray spectra of these regions have revealed that in most cases the brighter (and therefore denser) side of the edge is the colder side, and hence these jumps in gas density have been dubbed ``cold fronts'' (for a detailed review see \citep{MV07}).  The presence of one or more cold fronts is an indication of the motion of gas in a cluster. Famous examples of cold fronts formed by ongoing major mergers are those in Abell 3667 \citep{vik01} and the ``Bullet Cluster'' (1E0657-56) \citep{mar02}. However, cold fronts also occur in clusters which appear relaxed on the largest scales and show no obvious indications of major merging (e.g. Abell 1795 and Abell 2029). These fronts can arise due to ``sloshing'' motion of the core gas in the dark-matter dominated gravitational potential. This process was studied in detailed hydrodynamic simulations in \citet{asc06} (hereafter AM06), \citet{tit05} (though with a different interpretation), \citet{zuh10} (hereafter ZMJ10), and \citet{rod10a}. In these studies, the sloshing arises from encounters with small infalling groups or subclusters that gravitationally perturb the cluster core. Since the gas core is subject to ram pressure but the dark matter core is not, such perturbations can result in a separation between the gas and dark matter peaks, and consequently the core gas will begin to ``slosh'' back and forth in the gravitational potential well.

In ZMJ10, who used high-resolution grid-based simulations, it was demonstrated that if the ICM is modeled as an unmagnetized, inviscid fluid, the initially smooth and sharp cold fronts that result from sloshing motions are quickly disrupted by Kelvin-Helmholtz instabilities caused by the large shearing velocities present across the cold fronts. Such instabilities make the fronts appear ragged and torn, in contrast to the fronts that are seen in X-ray observations, which appear smooth and sharp. Additionally, mixing in these simulations was very efficient, resulting in the heating of the initially cool gas in the core as it was mixed with hotter gas brought into contact by sloshing. Conversely, ZMJ10 also showed that if the ICM is viscous, the instabilities are damped out and the resulting cold fronts retain their smooth, sharp shape. The smoothness of observed cold fronts in clusters of galaxies indicates that the ICM might possess a significant viscosity. An additional effect of viscosity in the ZMJ10 simulations was to suppress mixing of gases of different entropies, preventing cluster cool cores from heating due to such mixing. 

An alternative mechanism for the stability of cold fronts, which has been discussed previously in the literature, is the effect of magnetic fields. Strong observational evidence points to the existence of magnetic fields permeating the cluster volume \citep[see][for recent reviews]{car02,fer08}. One such line of evidence is synchrotron radio emission from sources such as radio halos \citep{fer01,gov01}, radio mini-halos in the cluster cool cores \citep{bur92,bac03,ven07,git07,gov09,gia11}, and radio relics. It is possible to estimate the magnetic field of the ICM under the assumption of equipartition between the relativistic electron population and the magnetic field \citep{pac70}. The magnetic field can also be estimated by comparing the synchrotron radio emission with inverse Compton hard X-ray emission from the same relativistic electrons, without the need to assume equipartition between the particles and the field. However, nonthermal X-ray emission from clusters has so far eluded confident detection--early reports of excess over the cluster thermal emission at high energies (Coma, Fusco-Femiano et al.\ 2004; A2163, Rephaeli et al.\ 2006) either allow an alternative thermal explanation or contradict the more recent upper limits from the Suzaku and SWIFT telescopes (e.g., Nakazawa et al. 2009; Sugawara et al. 2009; Ajello et al. 2009; Wik et al. 2009). Nondetection of inverse Compton emission corresponds to lower limits on the average magnetic field strengths of $0.2-0.5$ $\mu$G in radio halo regions (e.g., Ajello et al.\ 2009; Wik et al. 2009) and a few $\mu$G in relic regions (e.g., Finoguenov et al.\ 2010), broadly consistent with the equipartition radio estimates and Faraday rotation data.

A second line of evidence for magnetic fields in clusters is that of Faraday rotation of polarized emission of radio sources. Rotation measure (RM) studies indicate that magnetic field strengths in clusters are on the order of a few $\mu$G, with strengths up to tens of $\mu$G in cluster cool cores \citep{per91,tay93,fer95,fer99,tay02,tay06,tay07,bon10}. Additionally, in some clusters it is possible to derive RM maps, which are typically quite patchy, indicating that the coherence length of the cluster magnetic field is on the order of 10~kpc or less. High-resolution RM maps have been used to infer the cluster magnetic field power spectrum \citep{vog03,mur04,vog05,gov06,gui08,gov10}. These studies indicate that the magnetic field power spectrum is similar to a Kolmogorov type $(P_B(k) \propto k^{-5/3})$, depending on the assumed value for the coherence length of the field fluctuations. 

Magnetic fields oriented parallel to a shearing surface will suppress the growth of the Kelvin-Helmholtz instability \citep{lan60,cha61}. Whether or not the stability of cold fronts can be provided by magnetic fields depends on the strength of the field and the orientation of the field with respect to the front surface. What is required is a field oriented parallel to the front surface which has a magnetic pressure comparable to the kinetic energy per unit volume of the shearing flow across the front. For a cold front in the galaxy cluster A3667, \citet{vik01} and \citet{vik02} (hereafter V01 and V02) determined that the magnetic field strength required to stabilize the front is $B \sim 10 \mu$G, roughly an order of magnitude higher than the field strengths usually inferred from RM estimates and synchrotron diffuse radio emission outside of the cooling core regions. Additionally, the indications from RM maps that the magnetic field is tangled with a small coherence length seems to argue against the existence of a large-scale field that could drape a cold front. 

However, the region surrounding a cold front is not a typical place in a galaxy cluster. As argued in V01/V02, a cold front moving through the intracluster medium will cause the flow of the surrounding ICM to move around it, creating a shear flow. \citet{lyu06} pointed out that such flows around cold fronts and radio bubbles in galaxy clusters lead to ``magnetic draping.'' Provided the motion of the front is super-Alfvenic, a weak, tangled magnetic field will be stretched by this shear flow to produce a layer parallel to the the front surface. The magnetic field energy in this layer will be increased due to shear amplification, possibly strong enough to stabilize the front against Kelvin-Helmholtz instabilities \citep{kes10}. A number of previous simulation works \citep[e.g][]{asa04,asa07,dur07,dur08} have demonstrated the stabilizing effect of magnetic fields for cold fronts and AGN-blown bubbles in simplified situations, where cold ``blobs'' and hot bubbles propagated in simple, typically stratified atmospheres and different field geometries (e.g., uniform, tangled) were considered. \citet{tak08} simulated more realistic mergers with $N$-body/magnetohydrodynamics simulations, showing that magnetic fields wrapped around merger cold fronts (such as in the Bullet Cluster) and stabilized them against instabilities. Here, we extend these studies by simulating the formation of sloshing cold fronts produced by an encounter of a relaxed galaxy cluster with a small, infalling subcluster. Our aim is to determine for realistic magnetic field strengths and configurations whether or not the magnetic fields are amplified and drape the cold fronts to a sufficient extent to suppress instabilities that would otherwise grow and disrupt the smoothness of the fronts.
 
The interplay between gas sloshing motions and magnetic fields in galaxy clusters may have other interesting implications. Observations of clusters with radio mini-halos in their cool cores revealed a spatial association between sloshing cold fronts and the observed radio emission \citep{maz08}, suggesting that the radio emission originates from the effects of the sloshing motions. This may come from turbulence generated by the sloshing motions, the amplification of magnetic field strengths due to the associated shear flows \citep{kes10}, or a combination of both. Another interesting observable effect is the large, amplified magnetic field structures produced by sloshing may produce signatures in RM maps associated with the sloshing structures. Full 3D MHD simulations of sloshing such as ours may determine if such effects may be associated with cold fronts in relaxed clusters. 

This is a first paper in a series describing simulations of magnetized gas sloshing in galaxy cluster cores. The purposes of this work are to a) give technical details of the simulation setup, b) present results on the effect of sloshing of the gas and magnetic fields of the cluster, and c) serve as the groundwork for future studies which will explore additional physics and observable effects of sloshing. This paper is organized as follows: in Section 2 we describe the characteristics of the simulations and the code. In Section 3 we describe the characteristics of gas sloshing in our simulations in the absence and presence of magnetic fields, in particular with regard to the effect of sloshing on the fields and the effect of the fields on the structure of the cold fronts. In Section 4 we discuss the implications of these results. Finally, in Section 5 we summarize our results and discuss future developments of this work. Throughout this paper we assume a flat $\Lambda$CDM cosmology with $h = 0.7$, $\Omega_{\rm m} = 0.3$, and $\Omega_b = 0.02h^{-2}$. 

\section{Simulations\label{sec:sims}}

\subsection{Method\label{sec:method}}

In our simulations, we solve the ideal MHD equations. Written in conservation form in Gaussian units, they are:
\begin{eqnarray}
\frac{\partial{\rho}}{\partial{t}} + \nabla \cdot (\rho{\bf v})= 0 \\
\frac{\partial{(\rho{\bf v})}}{\partial{t}} + \nabla \cdot (\rho{\bf vv} - {\bf BB}) + \nabla{p} = \rho{\bf g} \\
\frac{\partial{E}}{\partial{t}} + \nabla \cdot [{\bf v}(E+p) - {\bf B}({\bf v \cdot B})] = \rho{\bf g \cdot v} \\
\frac{\partial{\bf B}}{\partial{t}} + \nabla \cdot ({\bf vB} - {\bf Bv}) = 0
\end{eqnarray}
where
\begin{eqnarray}
p = p_{\rm th} + \frac{B^2}{8\pi} \\
E = \frac{\rho{v^2}}{2} + \epsilon + \frac{B^2}{8\pi}
\end{eqnarray}
where $p_{\rm th}$ is the gas pressure, and $\epsilon$ is the gas internal energy per unit volume. For all our simulations, we assume an ideal gas equation of state with $\gamma = 5/3$. 

We performed our simulations using FLASH 3, a parallel hydrodynamics/$N$-body astrophysical simulation code developed at the Center for Astrophysical Thermonuclear Flashes at the University of Chicago \citep{fry00,dub09}. FLASH uses adaptive mesh refinement (AMR), a technique that places higher resolution elements of the grid only where they are needed. We are interested in capturing sharp ICM features like shocks and cold fronts accurately, as well as resolving the inner cores of the cluster dark matter halos. It is particularly important to be able to resolve the grid adequately in these regions. AMR allows us to do so without needing to have the whole grid at the same resolution. 

FLASH 3 solves the equations of magnetohydrodynamics using a directionally unsplit staggered mesh algorithm \citep[USM;][]{lee09}. The USM algorithm used in FLASH 3 is based on a finite-volume, high-order Godunov scheme combined with a constrained transport method (CT), which guarantees that the evolved magnetic field satisfies the divergence-free condition \citep{eva88}. In our simulations, the order of the USM algorithm corresponds to the Piecewise-Parabolic Method (PPM) of \citet{col84}, which is ideally suited for capturing shocks and contact discontinuties (such as the cold fronts that appear in our simulations). The USM solver in FLASH comes equipped with two methods for interpolating magnetic fields in a divergenceless manner from coarse to fine blocks on the AMR grid. The first is by simple injection of the magnetic field components from the coarser cells to the finer neighboring cells. The second is a higher-order method described in \citet{bal01}. We find that the choice of either prescription does not affect our conclusions. 

The gravitational potential on the grid is set up as the sum of two ``rigid bodies'' corresponding to the contributions to the potential from both clusters. This approach to the modeling the potential is used for simplicity and speed over solving the Poisson equation for the matter distribution, and is an adequate approximation for our purposes. It is the same approach that we used in ZMJ10 for the resolution test; it will be described in detail in Section \ref{sec:ICs}. 

\subsection{Initial Conditions\label{sec:ICs}}

Our initial conditions in these idealized simulations have been set up in a manner very similar to ZMJ10, with some differences which we elaborate on here.

For the cluster dark matter profile we have chosen a \citet{her90} profile:
\begin{equation}
\rho_{\rm DM}(r) = \frac{M_{\rm tot}}{2{\pi}a^3}\frac{1}{(r/a)(1+r/a)^3}
\end{equation}
where $M_{\rm tot}$ and $a$ are the mass and scale length of the DM halo. The Hernquist profile shares with the more commonly employed \citet[NFW]{NFW97} profile a ``cuspy'' inner radial dependence of the dark matter density, but results in simpler expressions for the mass, potential, and particle distribution functions. Because we are interested in the consequences of the interaction for only the central regions of the main cluster, the difference in the density dependence for large radii is unimportant. The corresponding gravitational potential has a particularly simple form:
\begin{equation}
\Phi_{\rm DM}(r) = -\frac{GM_{\rm tot}}{r+a}
\end{equation}
which for $r \gg a$ behaves as a point mass potential and for $r \ll a$ is approximately constant. For our simulations, the same profile shape is used for both the main cluster and the subcluster. 

\begin{figure}
\begin{center}
\plotone{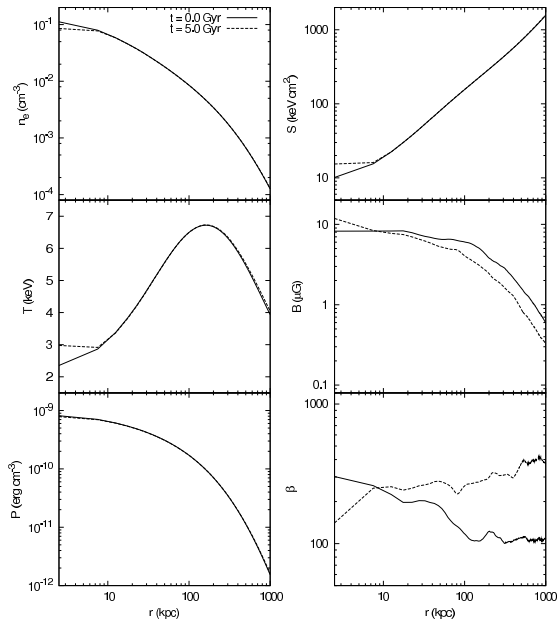}
\caption{Radial profiles of gas density, gas temperature, gas pressure, gas entropy, magnetic field strength, and plasma $\beta$ at the epochs $t$ = 0.0 Gyr and $t$ = 5.0 Gyr for a single cluster evolved in isolation.\label{fig:stability_test}}
\end{center}
\end{figure}

For the gas temperature, we use a phenomenological formula:
\begin{equation}
T(r) = \frac{T_0}{1+r/a}\frac{c+r/a_c}{1+r/a_c}
\end{equation}
where $0 < c < 1$ is a free parameter that characterizes the depth of the temperature drop in the cluster center and $a_c$ is the characteristic radius of that drop. This functional form can reproduce cluster temperature profiles of many observed relaxed galaxy clusters, which have a characteristic temperature drop in the center due to cooling. With this temperature profile, the corresponding gas density can be derived by imposing hydrostatic equilibrium:
\begin{equation}
\rho_{\rm gas}(r) = \rho_0\left(1+\frac{r}{a_c}\right)\left(1+\frac{r/a_c}{c}\right)^\alpha\left(1+\frac{r}{a}\right)^\beta,
\end{equation}
with exponents
\begin{equation}
\alpha \equiv -1-n\frac{c-1}{c-a/a_c}, \beta \equiv 1-n\frac{1-a/a_c}{c-a/a_c}.
\end{equation}
We set $n = 5$ in order to have a constant baryon fraction at large radii, and we compute the value of $\rho_0$ from the constraint $M_{\rm gas}/M_{\rm DM} = \Omega_{\rm gas}/\Omega_{\rm DM}$. This gas profile resembles those of most cool-core clusters since it continues to increase with decreasing radius and does not have a flat core. From these profiles we can derive a radial dependence for the cluster entropy profile:
\begin{equation}
S \equiv k_BTn_e^{-2/3} \propto \left(1+\frac{r}{a_c}\right)^{-5/3}\left(1+\frac{r/a_c}{c}\right)^{(3-2\alpha)/3}\left(1+\frac{r}{a}\right)^{-(2\beta+3)/3}
\end{equation}
which resembles a power-law $S(r) \propto r^{1.0-1.2}$ over most of the radial range, in line with observations of cool-core clusters \citep[e.g.,][]{don06}. The initial radial profiles for the main cluster are given in Figure \ref{fig:stability_test}.

Our merging clusters consist of a large, ``main'' cluster, and a small infalling subcluster. They are characterized by the mass ratio $R \equiv M_1/M_2$, where $M_1 = M_0R/(1+R)$ and $M_2 = M_0/(1+R)$ are the masses of the main cluster and the infalling satellite, respectively. The subcluster potential center starts at a distance $d$ from the main cluster center, and with an initial impact parameter $b$. For all of our simulations in this parameter study, we choose the same subcluster mass ratio ($R$ = 5), distance $d$ = 3~Mpc, and impact parameter $b$ = 500~kpc. To scale the initial profiles for the various mass ratios of the clusters, the combinations $M_i/a_i^3$, $c_i$, and $a_{c,i}/a_i$ are held constant. For the main cluster, we chose $a_1$ = 600~kpc, $c_1$ = 0.17, and $a_{c,1}$ = 60~kpc, to resemble mass, gas density, and temperature profiles typically observed in real galaxy clusters. In particular, our main cluster closely resembles A2029, a hot, massive, relaxed cluster with a cool core that exhibits a cold front \citep[e.g.,][]{vik05, vik06, asc06}. For all of the simulations, we set up the main cluster within a cubical computational domain of width $L = 2.4$~Mpc on a side, with a finest cell size on our AMR grid of $\Delta{x} = 2.34$~kpc (see Appendix B for the results of a resolution test). 

The initial cluster velocities are chosen so that the total kinetic energy of the system is set to a fraction 0 $\leq K \leq$ 1 of its potential energy, approximating the objects as point masses:
\begin{equation}
E \approx (K-1)\frac{GM_1M_2}{d} = (K-1)\frac{R}{(1+R)^2}\frac{GM_0^2}{d}
\end{equation} 
So the initial velocities in the reference frame of the center of mass are set to
\begin{equation}
v_1 = \frac{R\sqrt{2K}}{1+R}\sqrt{\frac{GM_0}{d}} ; v_2 = \frac{\sqrt{2K}}{1+R}\sqrt{\frac{GM_0}{d}}
\end{equation}
For the simulations presented in this work, we have set $K = 1/2$. 

Since both the gravitational potential of the main cluster and of the subcluster are modeled as rigid bodies, in order for the subcluster to fall into the main cluster and cause the resulting sloshing motions, it must be set up on a realistic trajectory. For this we choose to fix the center of the main cluster potential at the center of the domain, at rest for the duration of the simulation, and the center of mass of the subcluster is assumed to fall within this potential as a point mass. From the initial conditions at the beginning of the simulation the subcluster's position {\bf x} and velocity {\bf v} is integrated over a timestep $\Delta{t}$ using the variable-timestep leapfrog method often used for integration of particles in $N$-body simulations \citep{hoc88}:
\begin{eqnarray}
{\bf x}_i^1 &=& {\bf x}_i^0 + {\bf v}_i^0\Delta{t}^0 \\
{\bf v}_i^{1/2} &=& {\bf v}_i^0 + \frac{1}{2}{\bf a}_i^0\Delta{t}^0 \\
{\bf v}_i^{n+1/2} &=& {\bf v}_i^{n-1/2} + C_n{\bf a}_i^n + D_n{\bf a}_i^{n-1} \\
{\bf x}_i^{n+1} &=& {\bf x}_i^n + {\bf v}_i^{n+1/2}\Delta{t}^n
\end{eqnarray}
with the coefficients $C_n$ and $D_n$ given by
\begin{eqnarray}
C_n &=& \frac{1}{2}\Delta{t}^n + \frac{1}{3}\Delta{t}^{n-1} + \frac{1}{6}\left[\frac{({\Delta{t}^n})^2}{\Delta{t}^{n-1}}\right] \\
D_n &=& \frac{1}{6}\left[\Delta{t}^{n-1} - \frac{({\Delta{t}^n})^2}{\Delta{t}^{n-1}}\right]
\end{eqnarray}
where $n$ is the time index, $i$ is the spatial index, and {\bf a} is the particle's acceleration. By using time-centered velocities and stored accelerations, this method achieves second-order time accuracy. 

\begin{table*}[thdp]
\caption{Simulation Parameter Space\label{tab:Params}}
\begin{center}
\begin{tabular}{cccc}
\hline
\hline
Simulation & Initial $\beta$ & Initial Field Configuration & Initial $\lambda_0$ (kpc) \\
\hline
NoFields    & N/A   & N/A           & N/A \\ 
Beta100     & 100    & Tangled     & 43 \\
Beta400     & 400    & Tangled     & 43 \\
Beta1600   & 1600  & Tangled     & 43 \\
Beta6400   & 6400  & Tangled     & 43 \\
Tight         & 400    & Tangled     & 15 \\
Loose        & 400    & Tangled     & 120  \\
Tangential & 400    & Tangential & N/A \\
\hline
\end{tabular}
\end{center}
\end{table*}

The gravitational potential at all points on the grid is assumed to be the sum of the respective potentials of the main cluster and the subcluster. The gravitational acceleration is then computed by finite differencing the potential. Since the frame we have chosen (with the main cluster fixed at the center) is not an inertial frame, we must also compute the inertial acceleration on the main cluster from the subcluster and add it to the acceleration from gravity. 

In a real cluster merger, the subcluster will be tidally stripped of its dark matter, growing smaller in mass, and (if it is bound to the main cluster) each subsequent passage will induce a correspondingly weaker disturbance on the main cluster core, until the subcluster is fully absorbed into the main cluster. Additionally, the subcluster's orbit would be altered due to the effects of dynamical friction from the surrounding dark matter of the main cluster. In our simplified model, the subcluster is not stripped of its mass, and there is no dynamical friction, so it travels on a closed orbit which would eventually take it on the same trajectory past the main cluster. This would result in a second disturbance that would be equal in magnitude to the original passage and be highly disruptive. In order to study the effects of the resulting sloshing in isolation without the interference of subsequent crossings of the subcluster, the aforementioned trajectory is followed until the center of the subcluster reaches the opposite side of the box, at which point it is assumed to follow a constant-velocity trajectory for the remainder of the simulation. As was seen in ZMJ10, that used a more realistic $N$-body representation of the clusters' DM components, the majority of the effects on the main cluster from the subcluster are due to the first core passage, so the essential features of the encounter are still captured. 

This gravitational potential setup is used for computational simplicity, and is adequate for our qualitative study of hydrodynamic effects. We find that this procedure for computing the trajectory of the subclusters' orbit and the total acceleration on the gas reproduces well the general characteristics of the sloshing features seen in AM06 and ZMJ10. A forthcoming paper detailing quantitatively the similarities and differences between the self-gravitating and the rigid potential sloshing setups supports our use of the rigid-potential approximation \citep{rod11}.

Finally, it remains to set up the magnetic field of the cluster. For a realistic cluster magnetic field, it is important to satisfy a few basic conditions. The first is the magnitude of the field itself. Typical measurements from Faraday rotation and synchrotron radiation measurements suggest field strengths of of 1-10 $\mu$G. This implies that the plasma $\beta$ for the magnetic field ($\beta = p/p_B$, where $p_B = B^2/8\pi$) is high, with typical values in the range 100-1000 \citep[e.g., a field of 3~$\mu$G at $r$ = 150~kpc in Abell 2029 corresponds roughly to $\beta \approx 600$][]{vik06}. Secondly, it is important to satisfy the constraint that $\nabla \cdot {\bf B} = 0$. 

In order to have these two conditions simultaneously satisfied, we use the following procedure, as in \citet{rus07} and \citet{rus10}. A random magnetic field $\tilde{\bf B}({\bf k})$ is set up in $\bf{k}$-space on a uniform grid using independent normal random deviates for the real and imaginary components of the field. Thus, the components of the complex magnetic field in $\bf{k}$-space are set up such that
\begin{eqnarray}
\tilde{B_x}(\bf{k}) &=& B_1[N(u_1) + iN(u_2)] \\
\tilde{B_y}(\bf{k}) &=& B_2[N(u_3) + iN(u_4)] \\
\tilde{B_z}(\bf{k}) &=& B_3[N(u_5) + iN(u_6)] 
\end{eqnarray}
where $N(u)$ is a function of the uniformly distributed random variable $u$ that returns Gaussian-distributed random values, and the values $B_i$ are field amplitudes. We adopt a dependence of the magnetic field amplitude $B(k)$ on the wavenumber $|\bf{k}|$ similar to (but not the same as) \citet{rus07} and \citet{rus10}:
\begin{equation}
B(k) \propto k^{-11/6}{\rm exp}[-(k/k_0)^2]{\rm exp}[-k_1/k]
\end{equation}
where $k_0$ and $k_1$ control the exponential cutoff terms in the magnetic energy spectra. The cutoff at high wavenumber $k_0$ roughly corresponds to the coherence length of the magnetic field $k_0 = 2\pi/\lambda_0$ \citep[e.g., the scale of the observed patches in the RM maps, see][]{} and we vary this for a few of our simulations. The cutoff at low wavenumber $k_1 = 2\pi/\lambda_1$ roughly corresponds to $\lambda_1 \approx r_{500}/2 \approx$~500~kpc, which is held fixed for all of our simulations. This field spectrum corresponds to a Kolmogorov shape for the energy spectrum $(P_B(k) \propto k^{-5/3})$ with cutoffs at large and small linear scales. This field is then Fourier transformed to yield ${\bf B}({\bf x})$, which is rescaled to have an average value of $\sqrt{8\pi{p}/\beta}$ to yield a field that has a pressure that scales with the gas pressure, i.e. to have a spatially uniform $\beta$ for the initial field. For our simulations, we try different values of $\beta$ to determine the effects of different initial field strengths, and different values of $\lambda_0$ to determine the effects of different initial field configuration. There are also predictions from simulations with magnetic fields and anisotropic heat conduction that in the cluster cool cores, the magnetic field orientation may be preferentially tangential due to the heat-flux bouyancy instability \citep[HBI, see, e.g.,][]{par08,bog09,par09}. For this reason, it would be interesting to examine the effect of an initial field configuration where the field lines were preferentially tangential. For this purpose, we include a simulation that has a purely tangential initial field similar to that used in \citet{bog09}, described in spherical coordinates by
\begin{eqnarray}
B_r &=& 0 \\ \label{eqn:Br}
B_\theta &=& 2B_0\sin{\theta}\cos{2\phi} \\ \label{eqn:Btheta}
B_\phi &=& -B_0\sin{2\phi}\sin{2\theta} \label{eqn:Bphi}
\end{eqnarray}
Its magnitude was then scaled by the gas pressure. 

The entire set of simulations is summarized in Table \ref{tab:Params}. Finally, for all configurations, a field consistent with $\nabla \cdot {\bf B} = 0$ is generated by ``cleaning'' the field of divergence terms in Fourier space via
\begin{equation}
\tilde{\bf B}({\bf k}) \rightarrow (\bf{I}-\bf{\hat{k}\hat{k}})\tilde{\bf B}({\bf k})
\end{equation}
where $\hat{\bf{k}}$ is the unit vector in $\bf{k}$-space, and {\bf I} is the identity operator. Upon transformation back to real space, the magnetic field satisfies $\nabla \cdot {\bf B} = 0$. This field is then interpolated onto our AMR grid in such a way that the condition $\nabla \cdot {\bf B} = 0$ is maintained.

Finally, to test the robustness of our initial model for the main cluster, we perform a test run with the main cluster kept unperturbed, with an average magnetic field strength near $\beta$ = 100-200. Figure \ref{fig:stability_test} shows the profiles of gas density, gas temperature, and gas entropy at the beginning of the simulation and at a later epoch ($t$ = 5~Gyr after the beginning of the simulation), demonstrating the stability of these cluster quantities at all radii excepting the innermost couple of resolution elements (of width $\sim$~2~kpc) due to force smoothing (a known numerical effect due to the inability to resolve the gravitational force on scales smaller than the grid resolution). This change is small compared to the evolution observed in our simulations of clusters undergoing mergers. However, the magnetic field itself does evolve to a slightly different configuration, an effect we will discuss in Appendix A. 

\section{Results\label{sec:results}}

\subsection{Sloshing of the Cool Core in an Unmagnetized Test Case\label{sec:descript}}

First, we will briefly describe the sloshing process due to subcluster mergers as elucidated in Section 3 of AM06 and Section 3 of ZMJ10. This general description is applicable to all of our simulations, regardless of the details of the initial magnetic field or its presence or absence, but we will be referring in this description to our ``control'' simulation {\it NoFields}, in which there is no magnetic field in the cluster gas. 

\begin{figure}
\begin{center}
\plotone{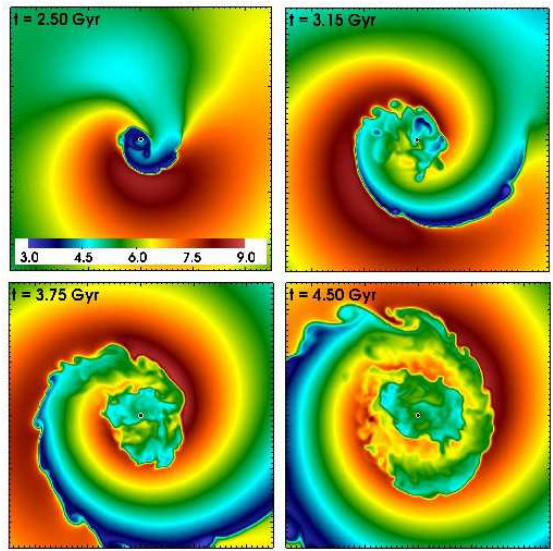}
\caption{Slices through the gas temperature for the simulation with no fields at various epochs. Each panel is 500 kpc on a side. Major tick marks indicate 100~kpc distances. The color scale shows temperature in keV. The black dot marks the position of the cluster potential minimum.\label{fig:nofields}}
\end{center}
\end{figure}

It is assumed that the subcluster has lost its gas due to ram pressure stripping from an earlier phase of the merger (although as the subcluster approaches the main cluster it begins to drag some of the cluster's ICM in a trailing sonic wake). This is seen in the first panel of Figure \ref{fig:nofields}, which shows slices of temperature through the cluster center in the plane of the clusters' mutual orbit. The core passage of the subcluster occurs at approximately $t \sim 1.8$~Gyr after the beginning of the simulation; each simulation is followed until $t = 5.0$~Gyr. As the subcluster approaches the main cluster's core and makes its passage, the gas and DM peaks of the main cluster feel the same gravity force toward the subcluster and move together towards it. However, the gas feels the effect of the ram pressure of the ambient medium. This fact becomes significant as the gas core is held back from the core of the dark matter by this pressure. After the passage of the core, when the direction of the gravitational force quickly changes, there is a rapid decline of the ram pressure. As a result from this change, the gas core experiences a ``ram pressure slingshot'' \citep{hal04}, where the gas that was previously held back by the ram pressure falls into the DM potential minimum and overshoots it. In addition to the gravitational disturbance, the wake trailing the subcluster transfers some of the angular momentum from the subcluster to the core gas and also acts to help push the core gas out of the DM potential well.

As the cool gas from the core climbs out of the potential minimum, it expands adiabatically. However, the densest, lowest-entropy gas quickly begins to sink back towards the potential minimum against the ram pressure from the surrounding ICM. Once again, as the cool gas falls into the potential well it overshoots it, and the process repeats itself on a smaller linear scale. Each time, a contact discontinuity (``cold front'') is produced. Due to the angular momentum transferred from the subcluster by the wake, these fronts have a spiral-shaped structure. Throughout this process, higher-entropy gas from larger radii is brought into contact with the lower entropy gas from the core, and as these gases mix, the entropy of the core gas is increased (ZMJ10). The last panel of Figure \ref{fig:nofields} shows that by the epoch $t = 4.5$~Gyr the initial temperature drop in the cluster center has disappeared. 

The resulting cold fronts in the simulation {\it NoFields} are qualitatively very similiar to the inviscid results from ZMJ10 (the only difference is the way the gravitational potential is modeled, which in that case was computed from the self-gravity of the DM and gas components of the cluster). The fronts start off very smooth, but quickly develop large Kelvin-Helmholtz-driven billows, due to the large shear flows that are present across the front surfaces. The action of the instabilities on the fronts is to make them appear jagged, and eventually even disappear (instead of the smooth fronts seen in observations). 

\subsection{The Effect of Sloshing on the Magnetic Field\label{sec:bfield_amp}}

We explore our parameter space by varying two conditions of the magnetic field: the initial average strength of the magnetic field and the initial spatial configuration of the magnetic field. We will discuss the evolution of the magnetic field in these two sets of simulations separately.

\begin{figure*}
\begin{center}
\plotone{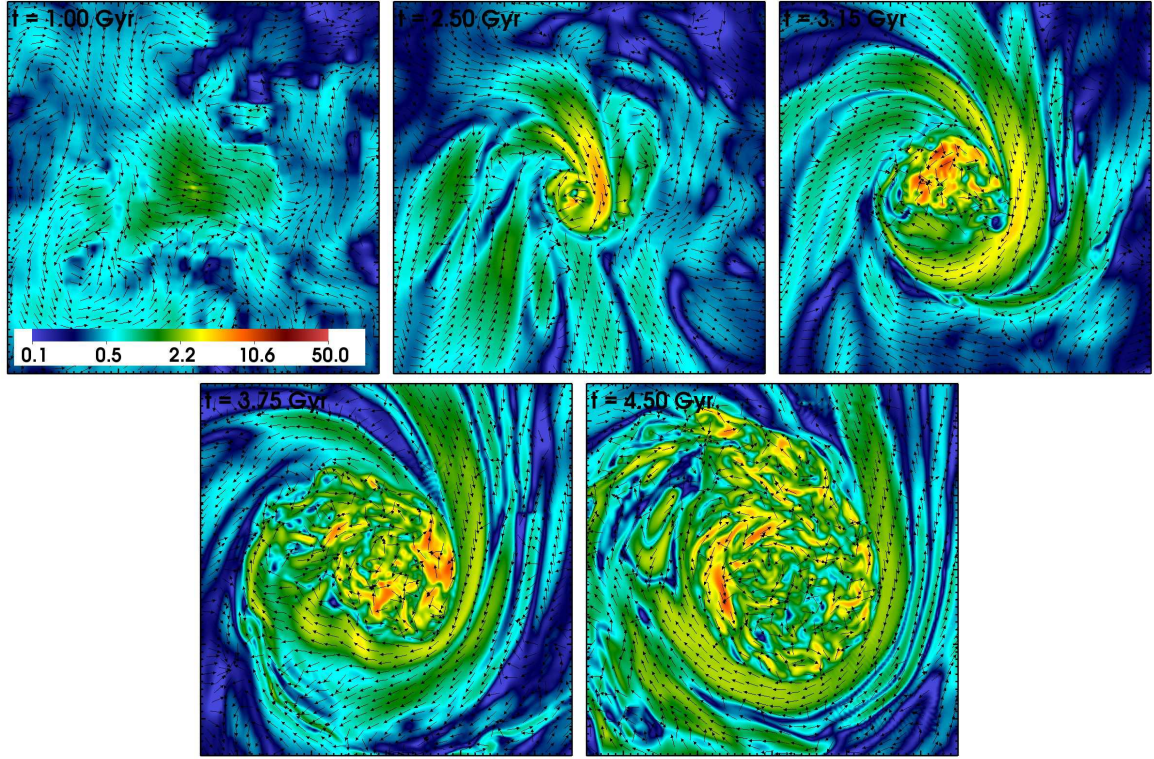}
\caption{Slices through the magnetic field strength with vectors indicating field direction in the x-y plane for the {\it Beta6400} simulation. Each panel is 500 kpc on a side. Major tick marks indicate 100~kpc distances. The color scale shows magnetic field strength in $\mu$G.\label{fig:bmag_beta6400}}
\end{center}
\end{figure*}

\begin{figure*}
\begin{center}
\plotone{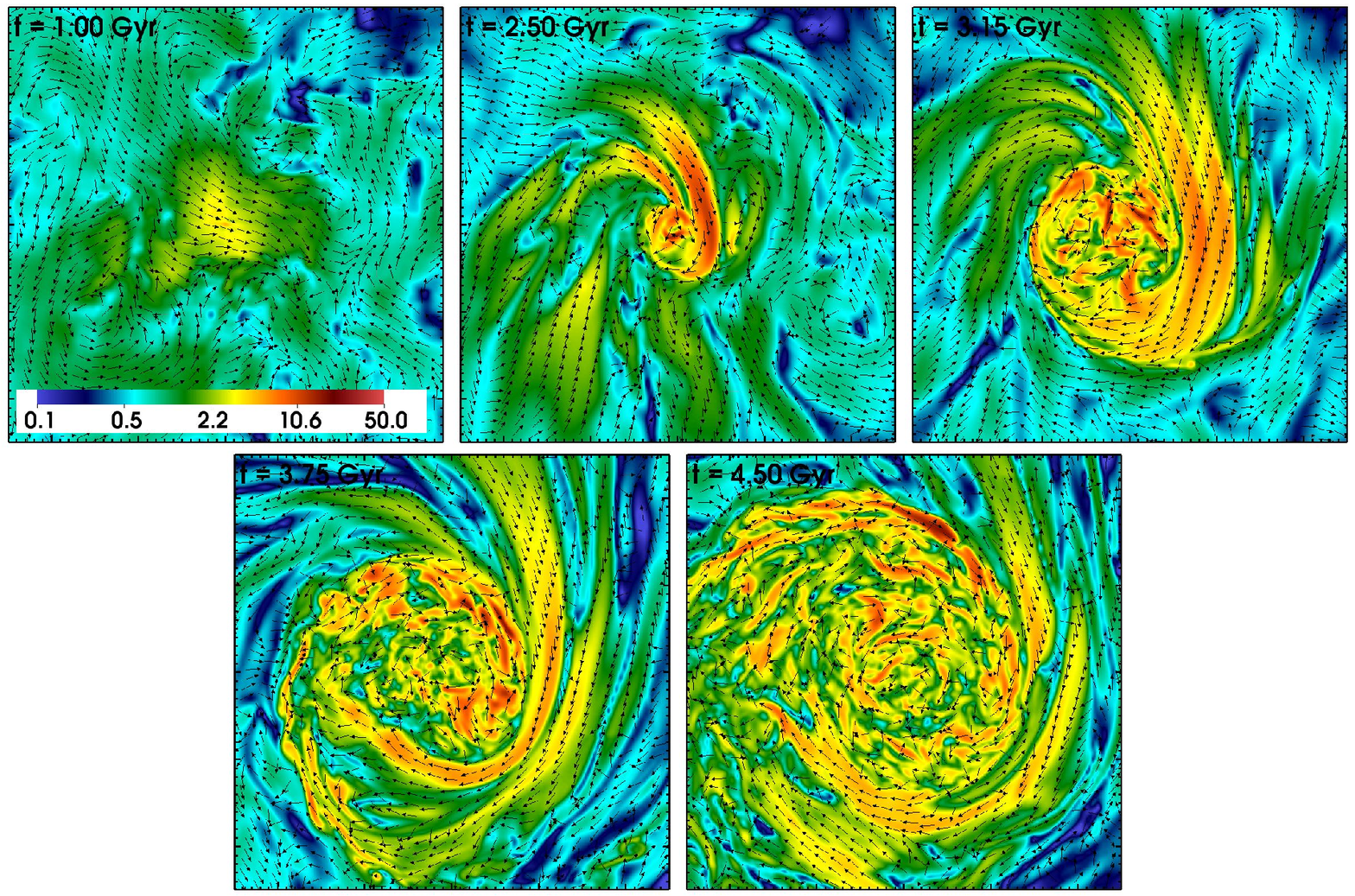}
\caption{Slices through the magnetic field strength with vectors indicating field direction in the x-y plane for the {\it Beta1600} simulation. Each panel is 500 kpc on a side. Major tick marks indicate 100~kpc distances. The color scale shows magnetic field strength in $\mu$G.\label{fig:bmag_beta1600}}
\end{center}
\end{figure*}

\begin{figure*}
\begin{center}
\plotone{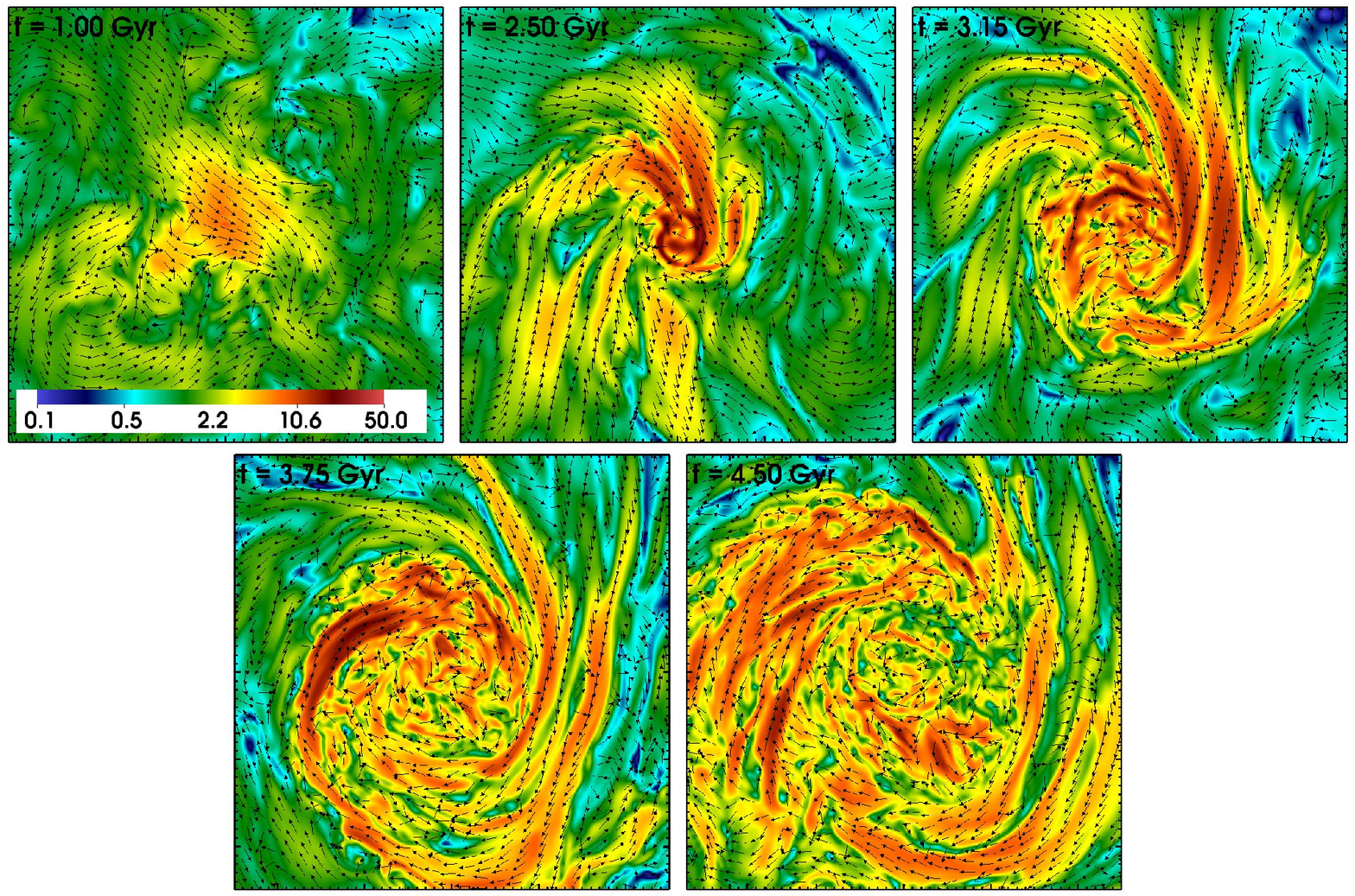}
\caption{Slices through the magnetic field strength with vectors indicating field direction in the x-y plane for the {\it Beta400} simulation. Each panel is 500 kpc on a side. Major tick marks indicate 100~kpc distances. The color scale shows magnetic field strength in $\mu$G.\label{fig:bmag_beta400}}
\end{center}
\end{figure*}

\begin{figure*}
\begin{center}
\plotone{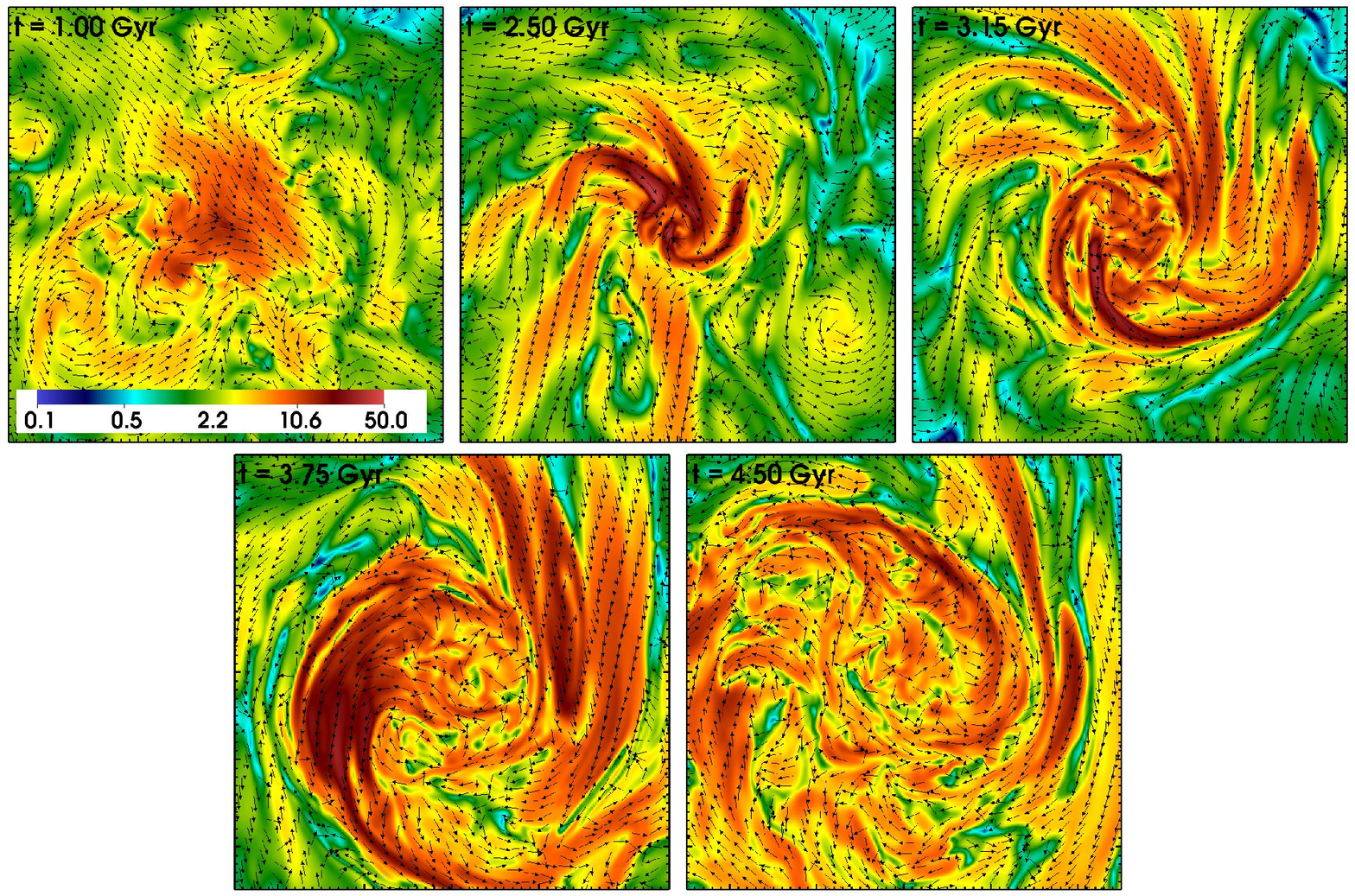}
\caption{Slices through the magnetic field strength with vectors indicating field direction in the x-y plane for the {\it Beta100} simulation. Each panel is 500 kpc on a side. Major tick marks indicate 100~kpc distances. The color scale shows magnetic field strength in $\mu$G.\label{fig:bmag_beta100}}
\end{center}
\end{figure*}

\begin{figure*}
\begin{center}
\plotone{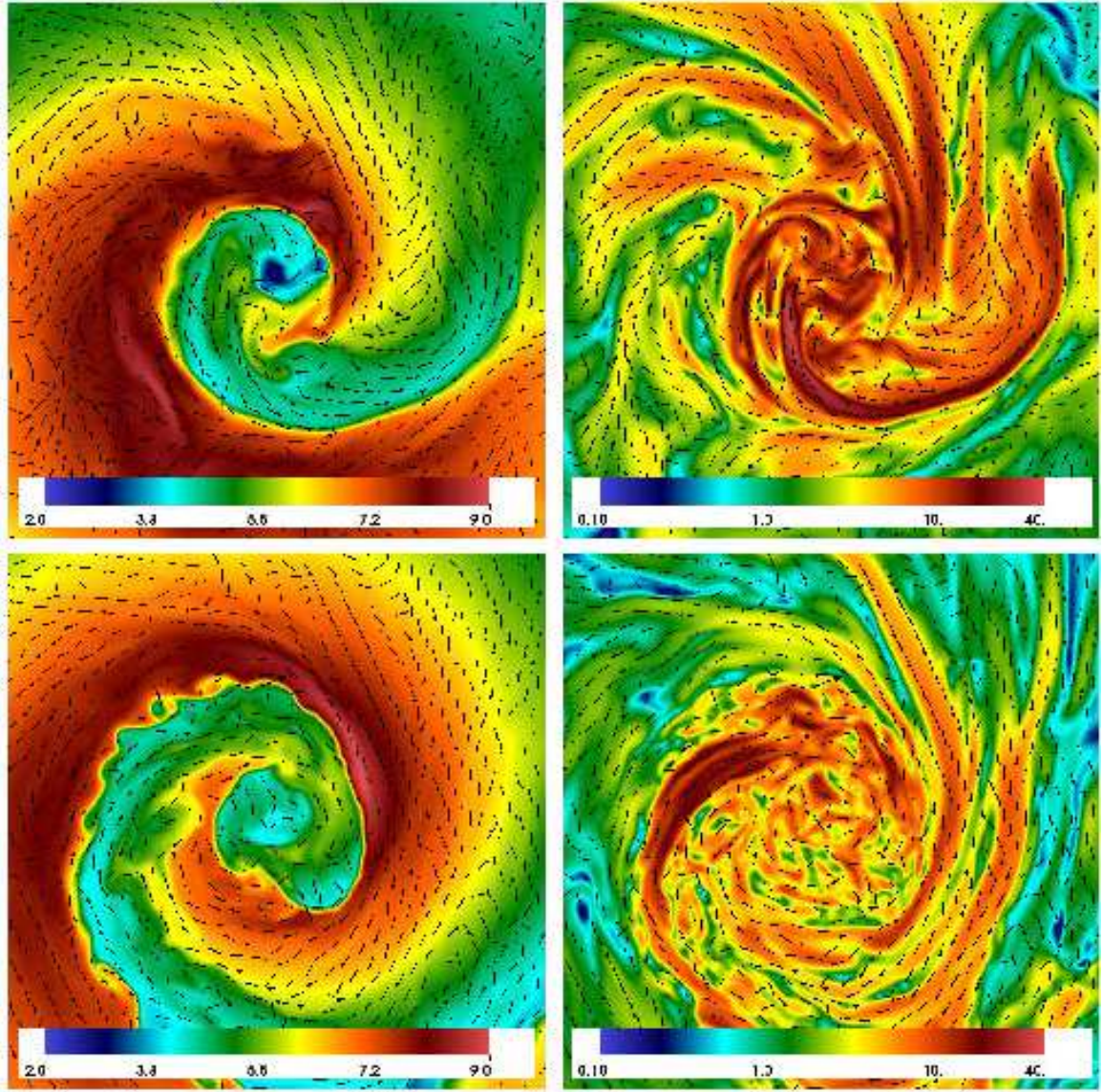}
\caption{The alignment of cold fronts with strongly magnetized layers. Top panels: Temperature (keV, left) and magnetic field strength ($\mu$G, right) for the {\it Beta100} simulation at the epoch $t$ = 3.15~Gyr. Bottom panels: Temperature (keV, left) and magnetic field strength ($\mu$G, right) for the {\it Beta400} simulation at the epoch $t$ = 3.75~Gyr. Vectors indicate magnetic field direction in the x-y plane. Each panel is 500 kpc on a side.\label{fig:fronts_alignment}}
\end{center}
\end{figure*}

\begin{figure}
\begin{center}
\plotone{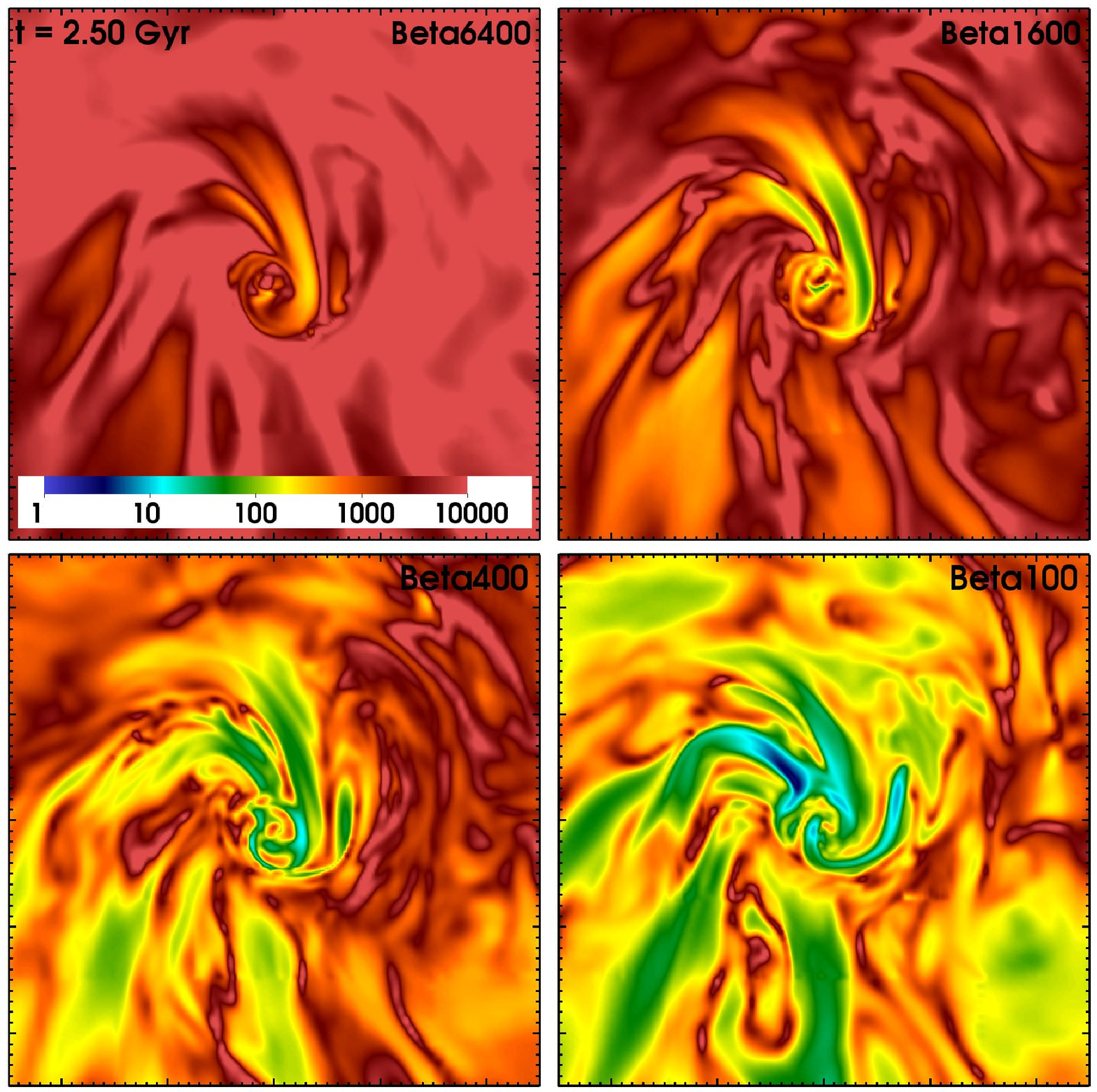}
\caption{Slices through the plasma $\beta$ for the simulations with varying initial $\beta$ at the epoch $t$ = 2.5 Gyr. Each panel is 500 kpc on a side. Major tick marks indicate 100~kpc distances.\label{fig:t2.5_beta}}
\end{center}
\end{figure}

\begin{figure}
\begin{center}
\plotone{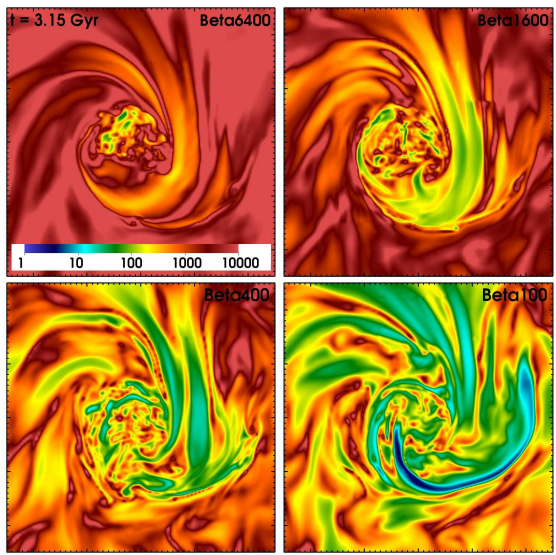}
\caption{Slices through the plasma $\beta$ for the simulations with varying initial $\beta$ at the epoch $t$ = 3.15 Gyr. Each panel is 500 kpc on a side. Major tick marks indicate 100~kpc distances.\label{fig:t3.15_beta}}
\end{center}
\end{figure}

All of the simulations begin with a roughly uniform $\beta$ with small fluctuations. Since the measured field strengths in clusters of galaxies is uncertain by an order of magnitude or so, it is important to characterize the effect of varying the initial strength of the magnetic field on its subsequent evolution. The simulation set \{{\it{Beta100,Beta400,Beta1600,Beta6400}}\} explores the effects of varying the initial magnetic field strength for a range of field strengths that covers the currently observationally plausible interval. The initial magnetic field configuration for each of these simulations is the same, corresponding to a Kolmogorov power spectrum with $k_0 = 2\pi/(43~{\rm kpc})$ and $k_1 = 2\pi/(500~{\rm kpc})$ (see Section \ref{sec:ICs}). 

The sloshing motions drag the field around, amplifying it along shear flows. Figures \ref{fig:bmag_beta6400} through \ref{fig:bmag_beta100} illustrate the effect of sloshing on the magnetic field strength and direction over the course of time. The first panel in each figure shows the tangled magnetic field before the sloshing motions begin, with the radially averaged field strength highest at the center and steadily decreasing outward. Over the course of the next few Gyr as represented in the following panels, the sloshing motions amplify the field and increase the size of the strongly magnetized region. These fields are also ordered on large scales, aligned largely along the cold front surfaces and other places in the domain where there are shear flows. In these layers the field strengths can be amplified up to tens of $\mu$G. Within the envelope of the cold fronts, at later epochs the fields are once again tangled, though the field strengths have been increased by a factor of $\sim$5-10 over their initial values. From these figures some inferences can be made about the physical reasonableness of the initial conditions of the magnetic field. In particular, the sloshing motions in the {\it Beta100} simulation result in magnetic field strengths that is over $\sim$10~$\mu$G out to radii of $r \sim$~150~kpc (Figure \ref{fig:bmag_beta100}), potentially in conflict with observations. However, in each simulation we do not find any magnetic field strengths in excess of tens of $\mu$G, implying that the fields will not increase without limit but will saturate.

\begin{figure}
\begin{center}
\plotone{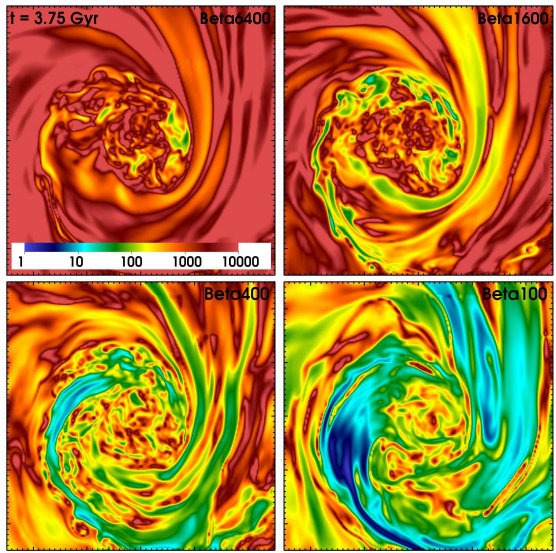}
\caption{Slices through the plasma $\beta$ for the simulations with varying initial $\beta$ at the epoch $t$ = 3.75 Gyr. Each panel is 500 kpc on a side. Major tick marks indicate 100~kpc distances.\label{fig:t3.75_beta}}
\end{center}
\end{figure}

\begin{figure}
\begin{center}
\plotone{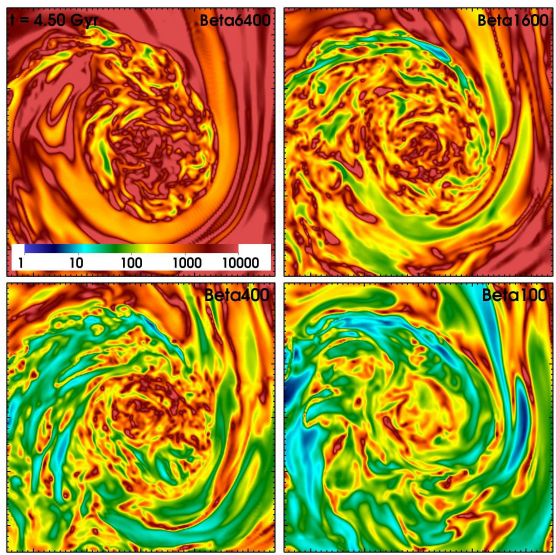}
\caption{Slices through the plasma $\beta$ for the simulations with varying initial $\beta$ at the epoch $t$ = 4.5 Gyr. Each panel is 500 kpc on a side. Major tick marks indicate 100~kpc distances.\label{fig:t4.5_beta}}
\end{center}
\end{figure}

Figure \ref{fig:fronts_alignment} shows side-by-side examples of the temperature and the magnetic field strength for two simulations at two different epochs (with magnetic field vectors overlaid), demonstrating the alignment of the strongly magnetized layers with the cold fronts. These layers are situated right underneath the cold front surfaces, and the field lines along these layers are stretched along the direction of the fronts, in agreement with the expectation from \citet{kes10}. 

\begin{figure}
\begin{center}
\plotone{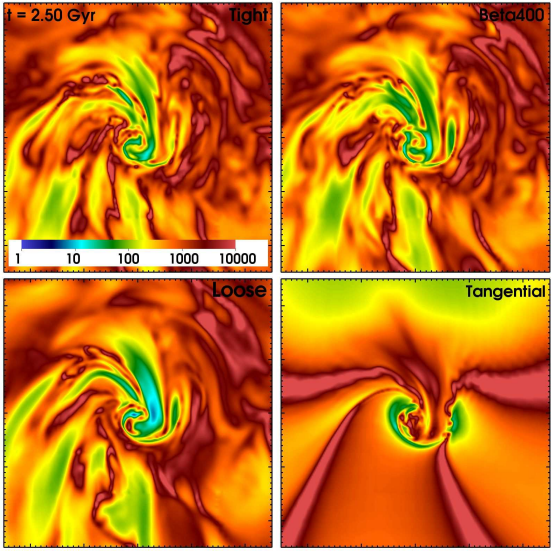}
\caption{Slices through the plasma $\beta$ for the simulations with varying initial configuration at the epoch $t$ = 2.5 Gyr. Each panel is 500 kpc on a side. Major tick marks indicate 100~kpc distances.\label{fig:t2.5_beta2}}
\end{center}
\end{figure}

\begin{figure}
\begin{center}
\plotone{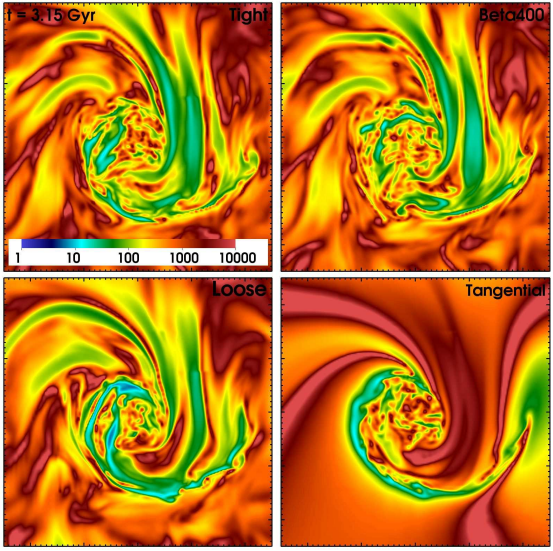}
\caption{Slices through the plasma $\beta$ for the simulations with varying initial configuration at the epoch $t$ = 3.15 Gyr. Each panel is 500 kpc on a side. Major tick marks indicate 100~kpc distances.\label{fig:3.15_beta2}}
\end{center}
\end{figure}

Figures \ref{fig:t2.5_beta} through \ref{fig:t4.5_beta} show slices of the plasma $\beta$ through the center of the domain for the epochs $t$ = 2.5, 3.15, 3.75, and 4.5~Gyr after the beginning of the simulation, for the simulations with varying initial $\beta$. In each case, the shear flows amplify the magnetic field, decreasing $\beta$. The degree to which $\beta$ is decreased in these layers is dependent on the initial $\beta$, however in each case the degree of amplification of the field energy is similar, on an order of magnitude or higher. 

\begin{figure}
\begin{center}
\plotone{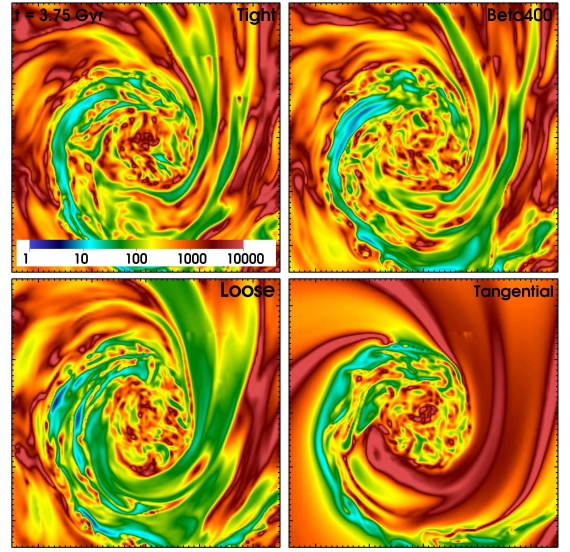}
\caption{Slices through the plasma $\beta$ for the simulations with varying initial configuration at the epoch $t$ = 3.75 Gyr. Each panel is 500 kpc on a side. Major tick marks indicate 100~kpc distances.\label{fig:t3.75_beta2}}
\end{center}
\end{figure}

\begin{figure}
\begin{center}
\plotone{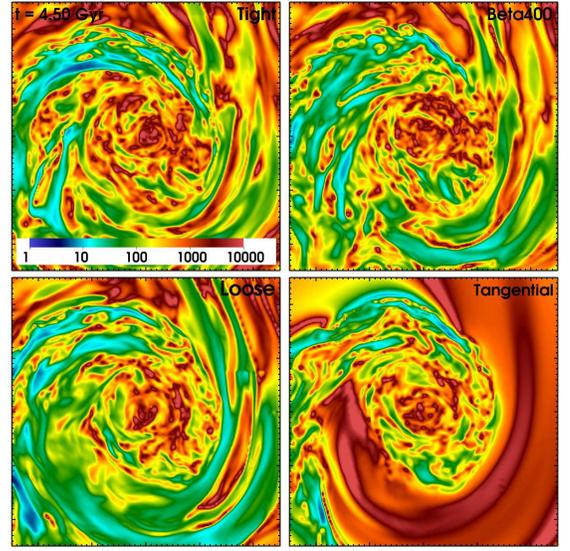}
\caption{Slices through the plasma $\beta$ for the simulations with varying initial configuration at the epoch $t$ = 4.5 Gyr. Each panel is 500 kpc on a side. Major tick marks indicate 100~kpc distances.\label{fig:t4.5_beta2}}
\end{center}
\end{figure}

Our default setup for the magnetic field spatial configuration, as detailed in Section \ref{sec:ICs}, is a tangled magnetic field. This field may be characterized by the slope of the initial power spectrum (which we take to be Kolmogrorov), the cutoff at large $k$ ($k_0$), and the cutoff at small $k$ ($k_1$). It is particularly instructive to examine the effect of varying the cutoff at large $k$ (small scales), since a magnetic field that is more tangled on smaller scales may be more difficult to amplify by shear amplification. The simulation set \{{\it{Beta400,Tight,Loose,Tangential}}\} explores the effects of varying the initial magnetic field spatial configuration, with ``Beta400'' being the default configuration with $k_0 = 2\pi/43~{\rm kpc}^{-1}$, ``Tight'' corresponding to a higher $k_0 = 2\pi/15~{\rm kpc}^{-1}$, ``Loose'' corresponding to a lower $k_0 =  2\pi/120~{\rm kpc}^{-1}$, and ``Tangential'' corresponding to a tangentially oriented field (Equations \ref{eqn:Br}-\ref{eqn:Bphi}). The initial average magnetic field strength for each of these simulations is the same, corresponding to $\beta$ = 400. Figures \ref{fig:t2.5_beta2} through \ref{fig:t4.5_beta2} show slices of the plasma $\beta$ through the center of the domain for the epochs $t$ = 2.5, 3.15, 3.75, and 4.5~Gyr after the beginning of the simulation. In contrast to our results for varying the initial $\beta$, there is not a similarly strong dependence of the evolution of the magnetic field on its initial configuration. About 2~Gyr after the initial disturbance, there is little qualitative difference between the runs. The field structure becomes similarly tangled and amplified to similar values.

\begin{figure*}
\begin{center}
\plotone{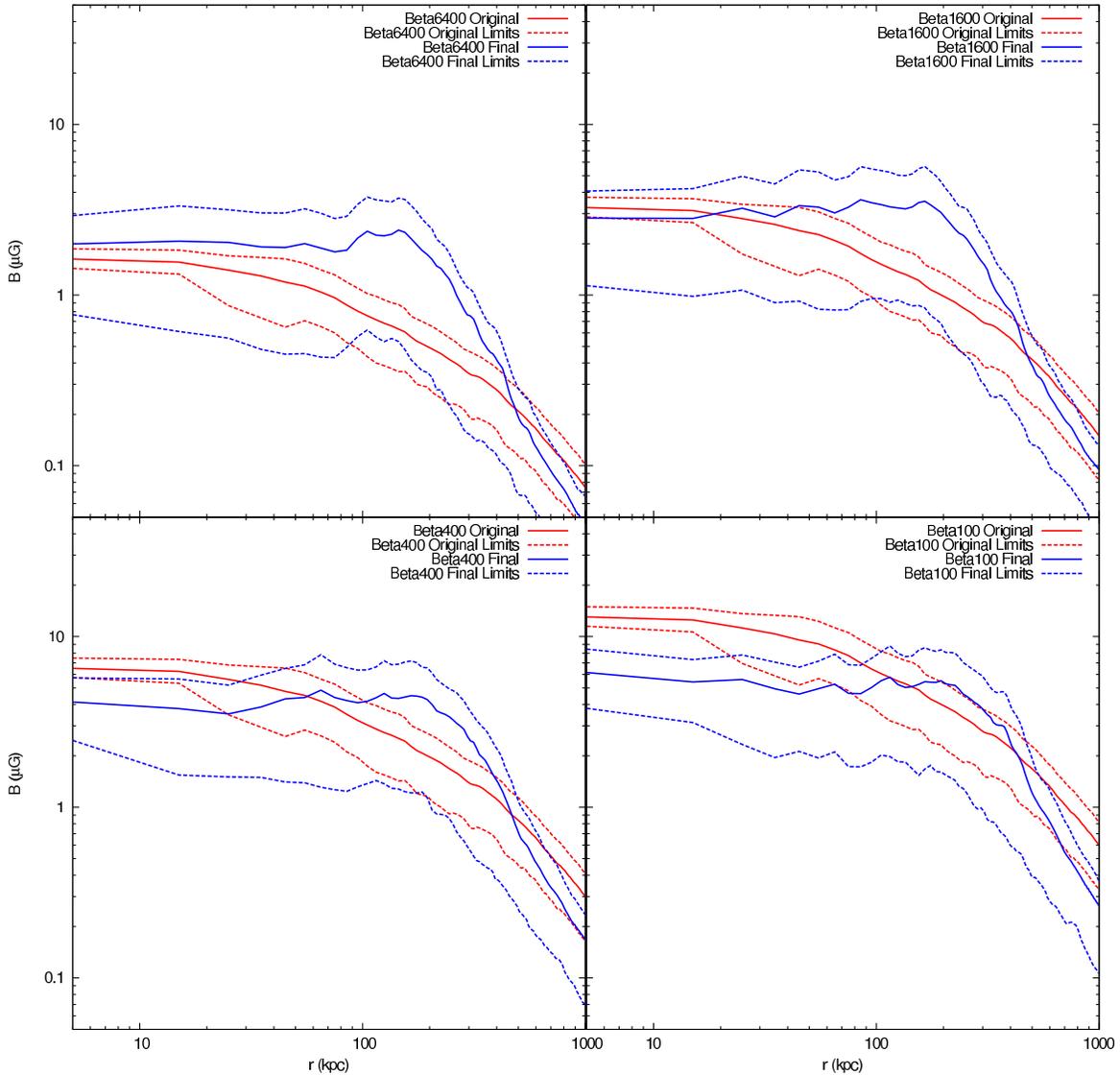}
\caption{Initial (red lines) and final (at $t$ = 5 Gyr, blue lines) profiles of the magnetic field strength for the simulations with varying initial $\beta$. Dot-dash lines indicate the final maximum and minimum values for each radial bin. Bin widths are 10~kpc.\label{fig:bmag_profiles}}
\end{center}
\end{figure*}

\begin{figure*}
\begin{center}
\plotone{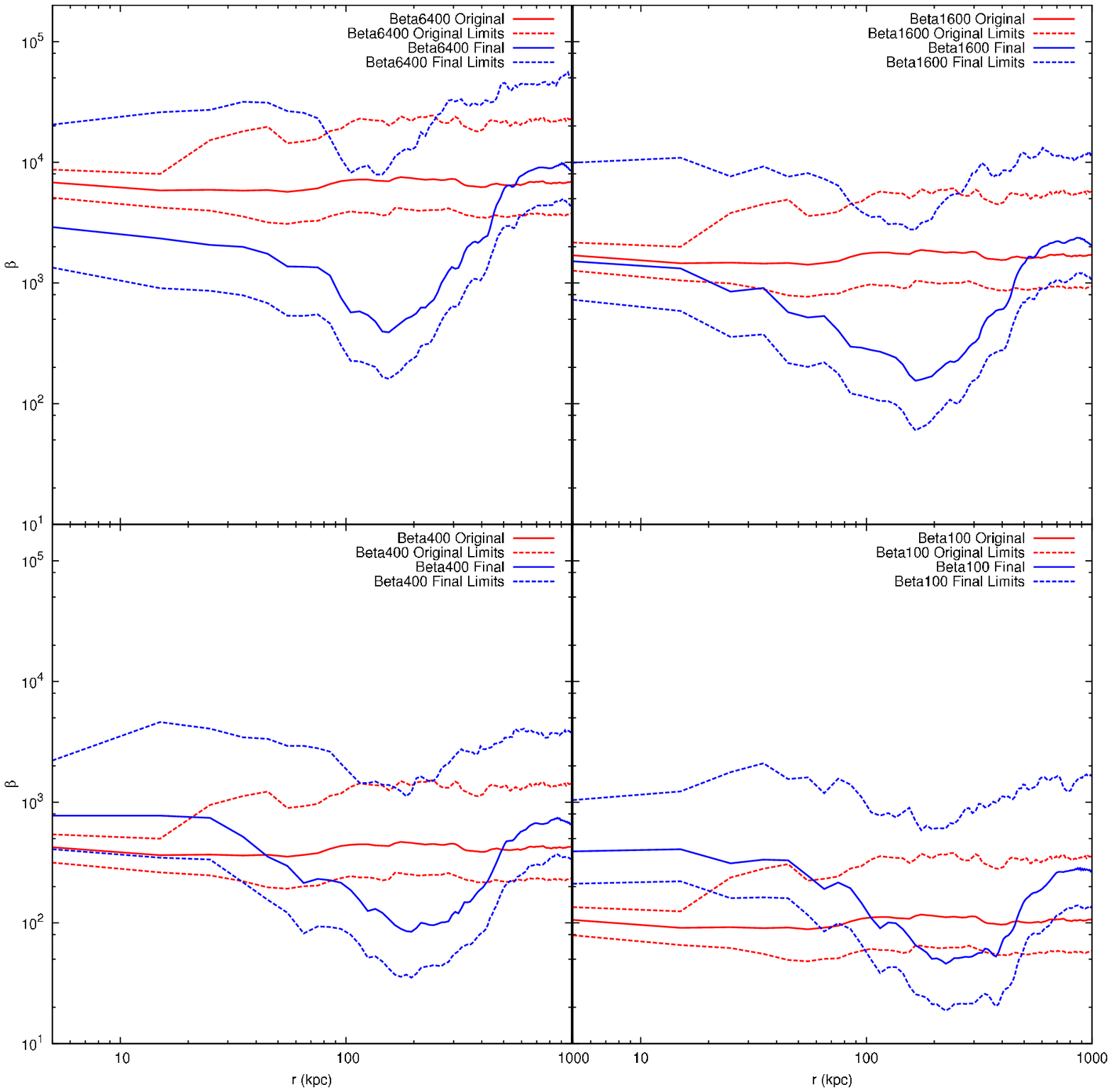}
\caption{Initial (red lines) and final (at $t$ = 5 Gyr, blue lines) profiles of the plasma $\beta$ for the simulations with varying initial $\beta$. Dot-dash lines indicate the final maximum and minimum values for each radial bin. Bin widths are 10~kpc.\label{fig:beta_profiles}}
\end{center}
\end{figure*}

Within the sloshing region, the magnetic field becomes extremely tangled with large fluctuations in strength. This is most apparent in the {\it Tangential} simulation, which begins with a relatively smooth tangentially oriented field but ends up with a randomly oriented field tangled on small scales within the sloshing region (see the last panels of Figures \ref{fig:t2.5_beta2}-\ref{fig:t4.5_beta2}). It appears that sloshing such as we are considering is more than enough to disrupt the tangential structure arising from the HBI, though future simulations including anisotropic heat conduction will be required to confirm this (ZuHone et al. 2011, in preparation). 

The field is most strongly amplified in the cold front layers confining the sloshing region, but what is the overall increase in magnetic energy due to this amplification? Figure \ref{fig:bmag_profiles} shows the radial profile of the magnetic field strength, and Figure \ref{fig:beta_profiles} shows the radial profile of the plasma $\beta$ of the gas, both at the beginning and the end of the simulations with varying initial $\beta$. The final profiles have a very similar shape across the simulations. Out to a radius of $r \sim 100$~kpc, the average magnetic field strength is roughly constant. There is an increase in field strength between radii $r \sim 100-500$~kpc (the radii of the cold fronts at this point in the simulations), and after this radius the field strength begins to decline with radius. The main difference between the profiles is their magnitude compared with the initial field strength. For the simulations with higher initial field strengths (simulations {\it Beta100} and {\it Beta400}), the final average field strength within $r \sim 100$~kpc is lower than the initial average field strength, though the field within the region of the cold fronts ($r \sim 100-500$~kpc) is still stronger. For the simulations with lower initial field strength (simulations {\it Beta1600} and {\it Beta6400}), we find that the average field is amplified at most radii from its initial value. For the simulations with varying initial spatial field configurations for $\beta = 400$, we find that the final profiles are very similar in magnitude and shape. 

Interestingly, the average field strengths for each simulation with varying initial $\beta$, and especially for simulations {\it Beta100} and {\it Beta400} (Figure \ref{fig:bmag_profiles} and Figure \ref{fig:beta_profiles}), are all similar within a factor of $\sim$3 (with corresponding field energies similar within an order of magnitude), despite the initial variation in field strength over an order of magnitude. Though sloshing increases magnetic field strengths at the cold fronts, that the final average magnetic energy within the sloshing region ends up being the same regardless of the initial field strength or structure (within reasonable limits, corresponding to $B \sim {\rm a~few}~\mu$G or $\beta \sim$~several hundred).

\begin{figure}
\begin{center}
\plotone{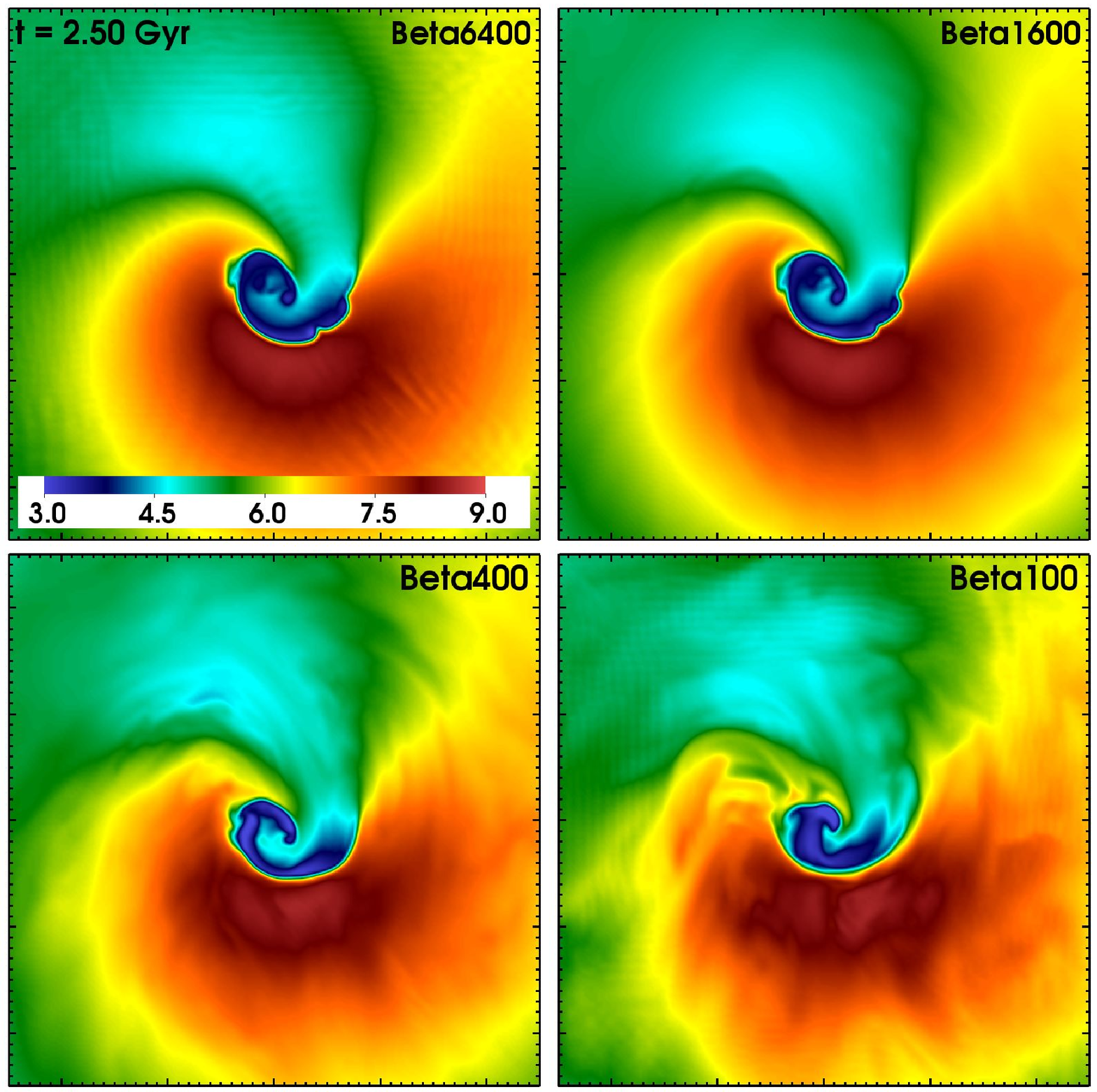}
\caption{Slices through the gas temperature for the simulations with varying initial $\beta$ at the epoch $t$ = 2.5 Gyr. Each panel is 500 kpc on a side. Major tick marks indicate 100~kpc distances. The color scale shows temperature in keV.\label{fig:t2.5_temp}}
\end{center}
\end{figure}

\begin{figure}
\begin{center}
\plotone{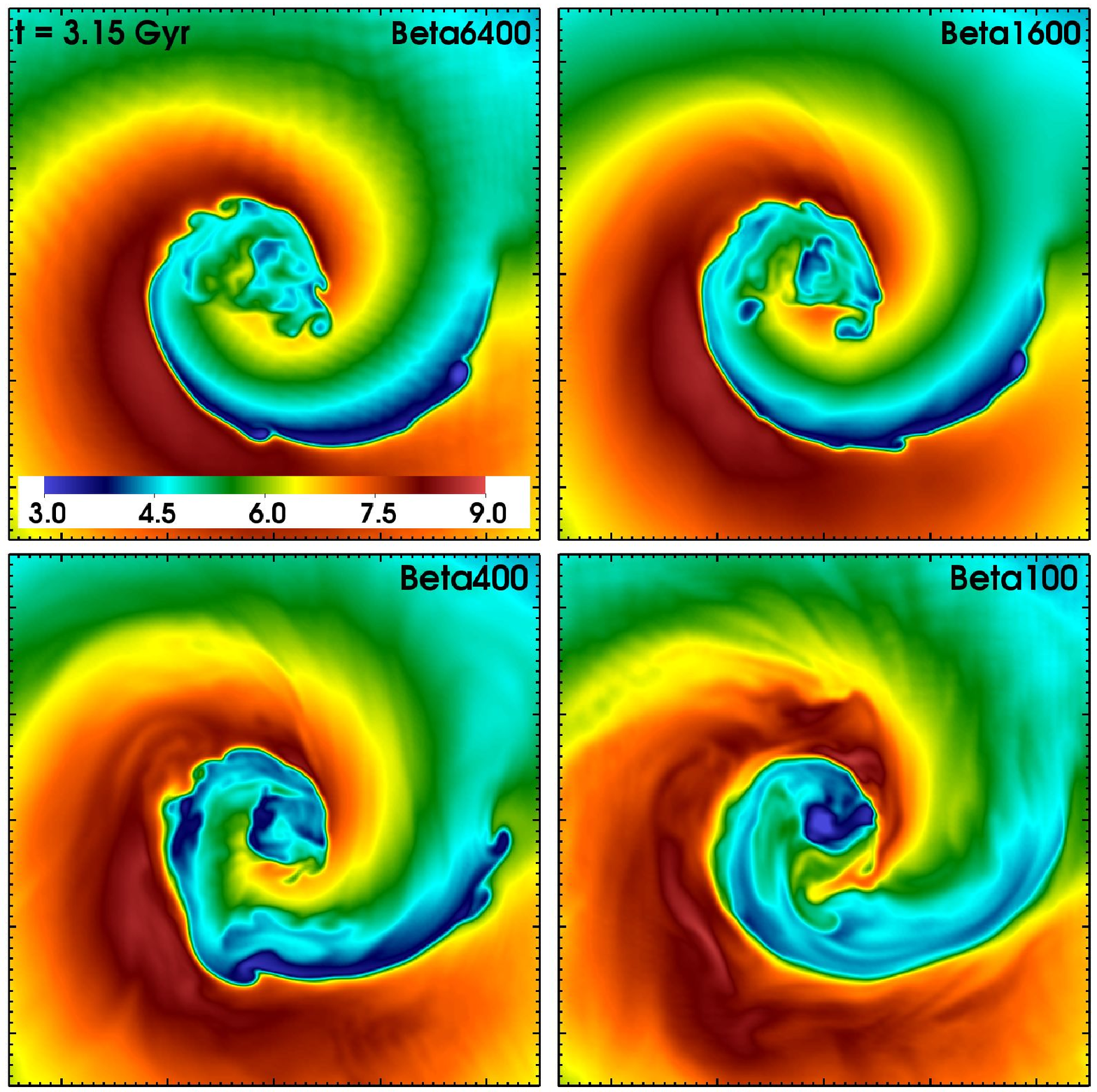}
\caption{Slices through the gas temperature for the simulations with varying initial $\beta$ at the epoch $t$ = 3.15 Gyr. Each panel is 500 kpc on a side. Major tick marks indicate 100~kpc distances. The color scale shows temperature in keV.\label{fig:t3.15_temp}}
\end{center}
\end{figure}

\begin{figure}
\begin{center}
\plotone{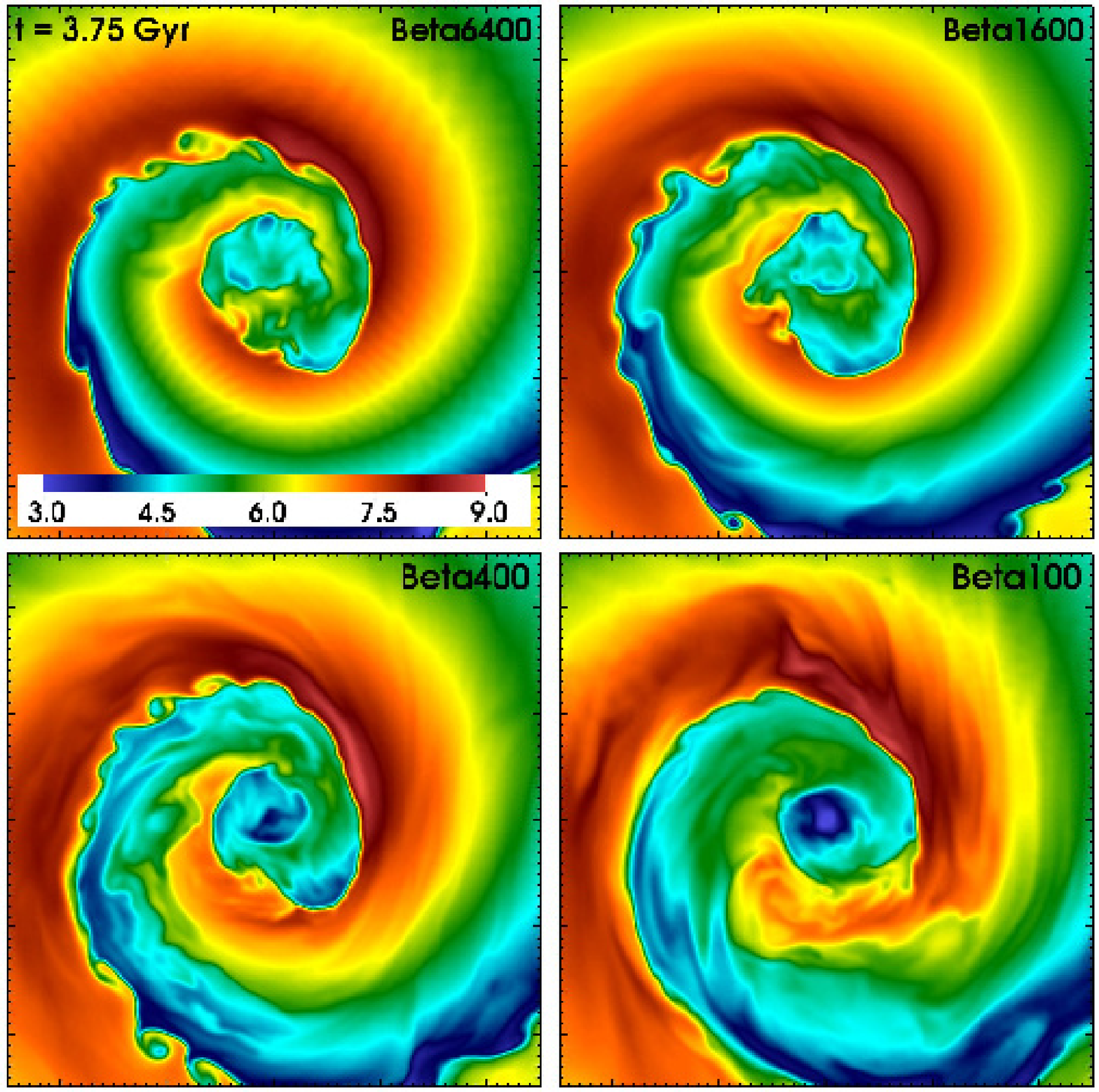}
\caption{Slices through the gas temperature for the simulations with varying initial $\beta$ at the epoch $t$ = 3.75 Gyr. Each panel is 500 kpc on a side. Major tick marks indicate 100~kpc distances. The color scale shows temperature in keV.\label{fig:t3.75_temp}}
\end{center}
\end{figure}

\begin{figure}
\begin{center}
\plotone{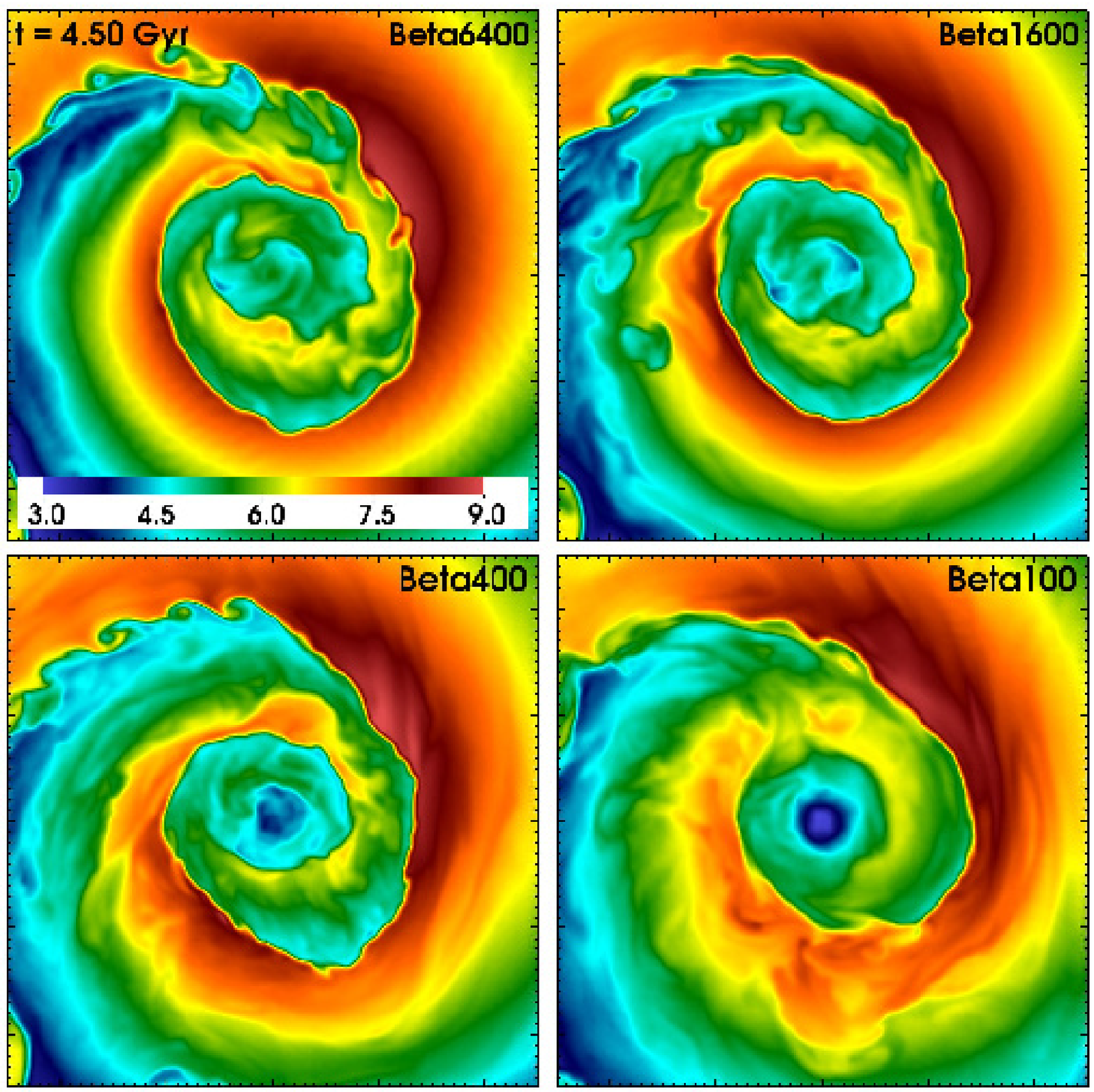}
\caption{Slices through the gas temperature for the simulations with varying initial $\beta$ at the epoch $t$ = 4.5 Gyr. Each panel is 500 kpc on a side. Major tick marks indicate 100~kpc distances. The color scale shows temperature in keV.\label{fig:t4.5_temp}}
\end{center}
\end{figure}

\begin{figure}
\begin{center}
\plotone{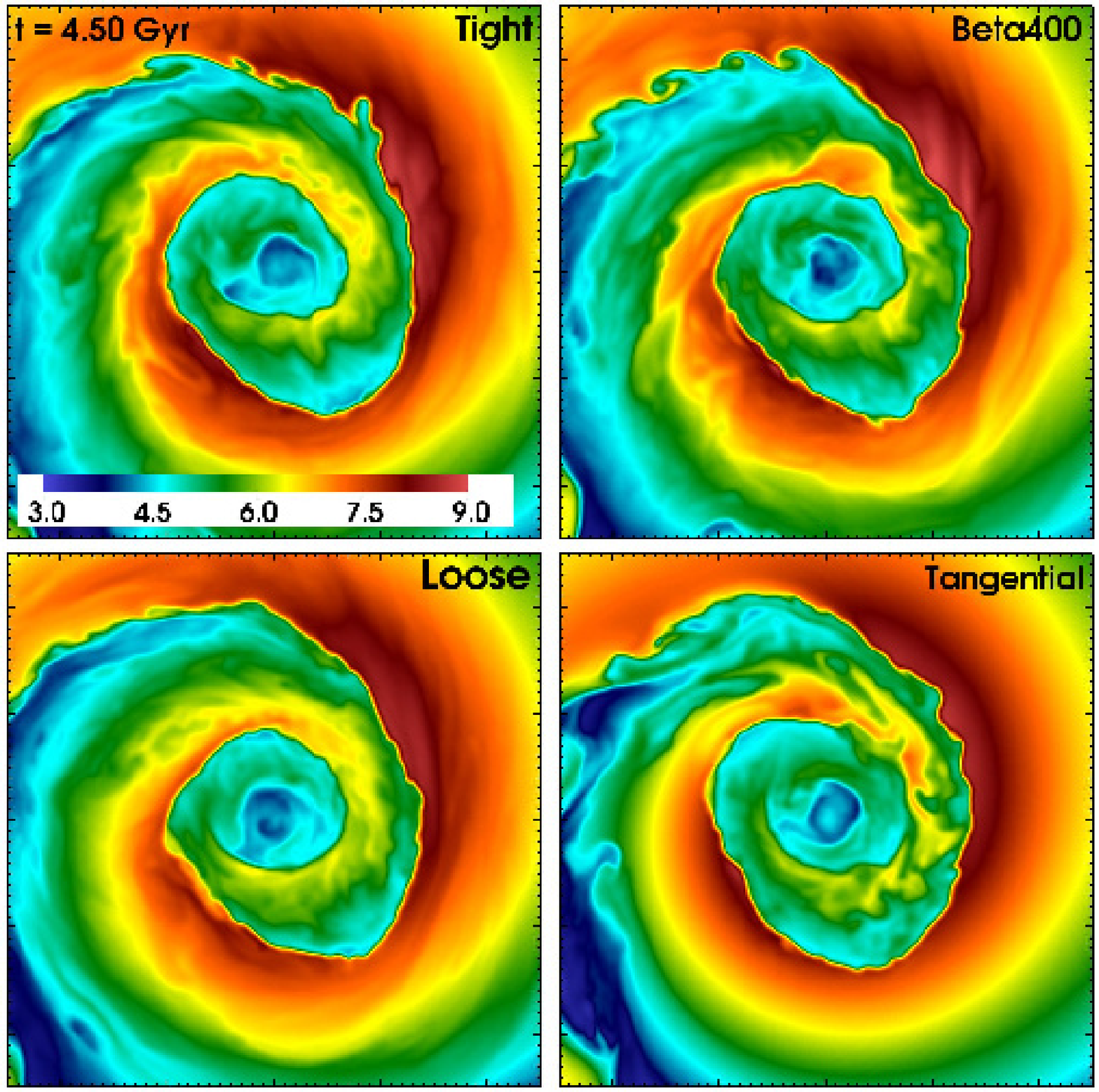}
\caption{Slices through the gas temperature for the simulations with varying initial configuration at the epoch $t$ = 4.5 Gyr. Each panel is 500 kpc on a side. Major tick marks indicate 100~kpc distances. The color scale shows temperature in keV.\label{fig:t4.5_temp2}}
\end{center}
\end{figure}

\begin{figure*}
\begin{center}
\includegraphics[width=0.9\textwidth]{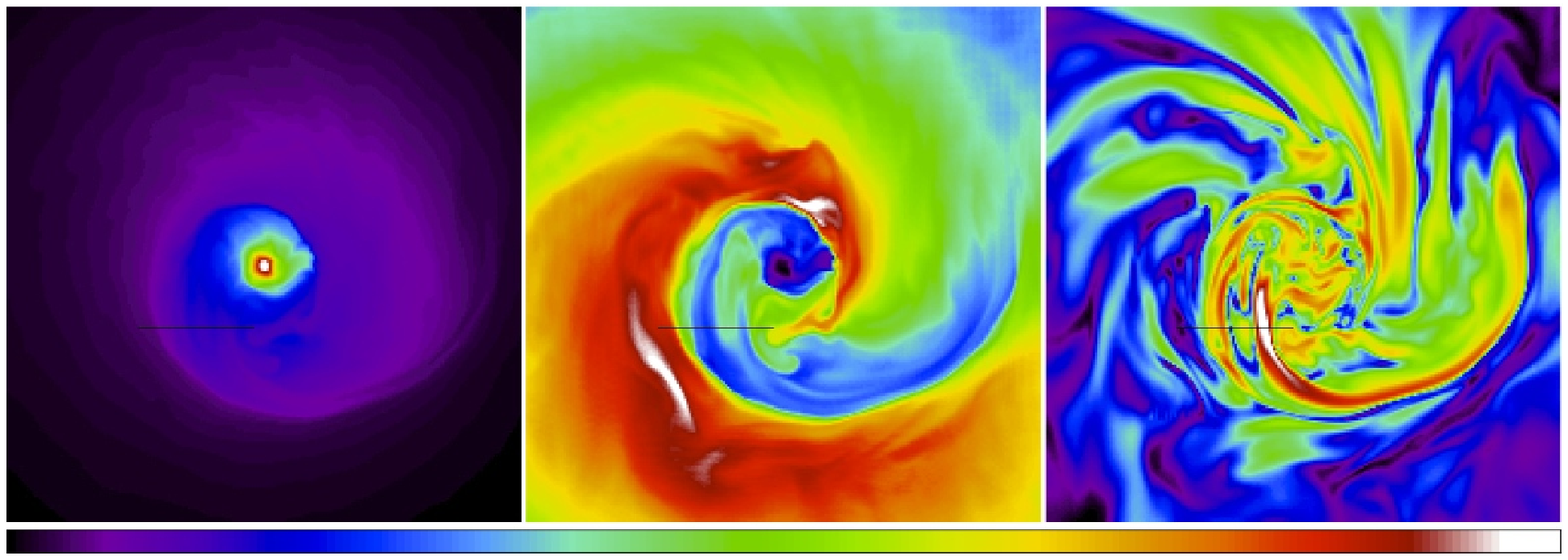}
\par
\includegraphics[width=0.9\textwidth]{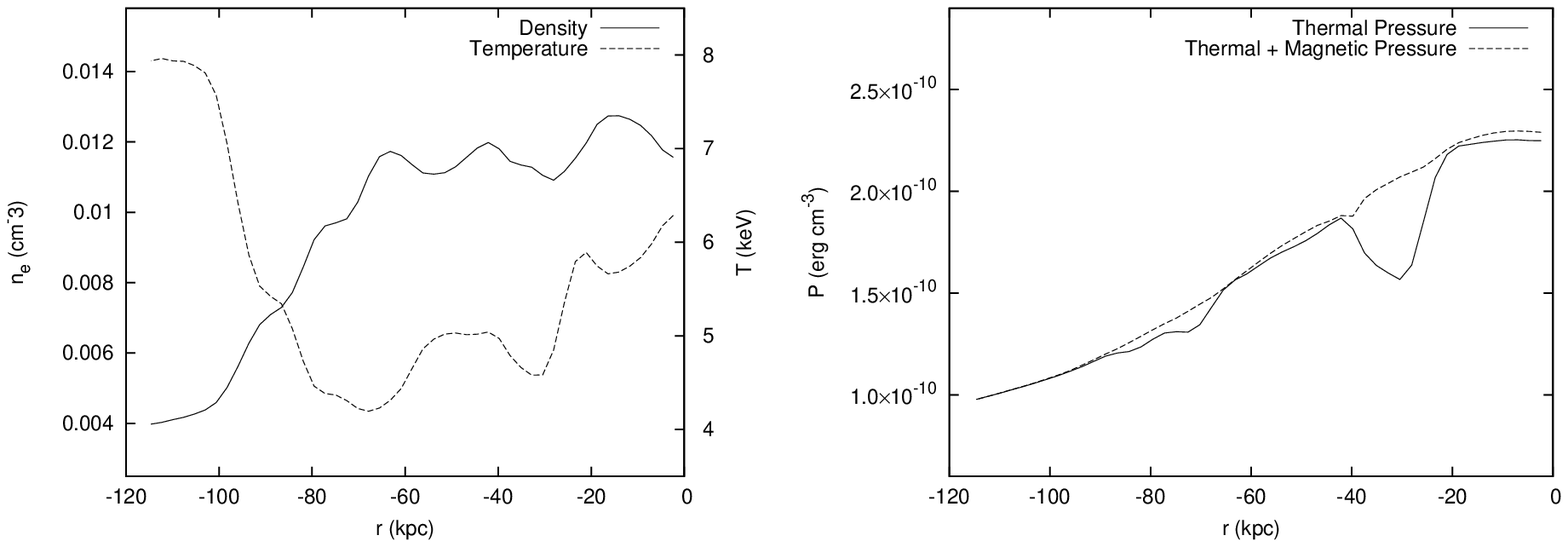}
\caption{An example of temperature and density fluctuations as a result of magnetic pressure from the {\it Beta100} simulation at the epoch $t$ = 3.15 Gyr. Top panels: Gas density, temperature, and magnetic pressure slices through the center of the domain. Bottom panels: Profiles of the density, temperature, thermal pressure, and total (magnetic + thermal) pressure along the black lines in the top panels.\label{fig:fluctuations}}
\end{center}
\end{figure*}

\begin{figure*}
\begin{center}
\plottwo{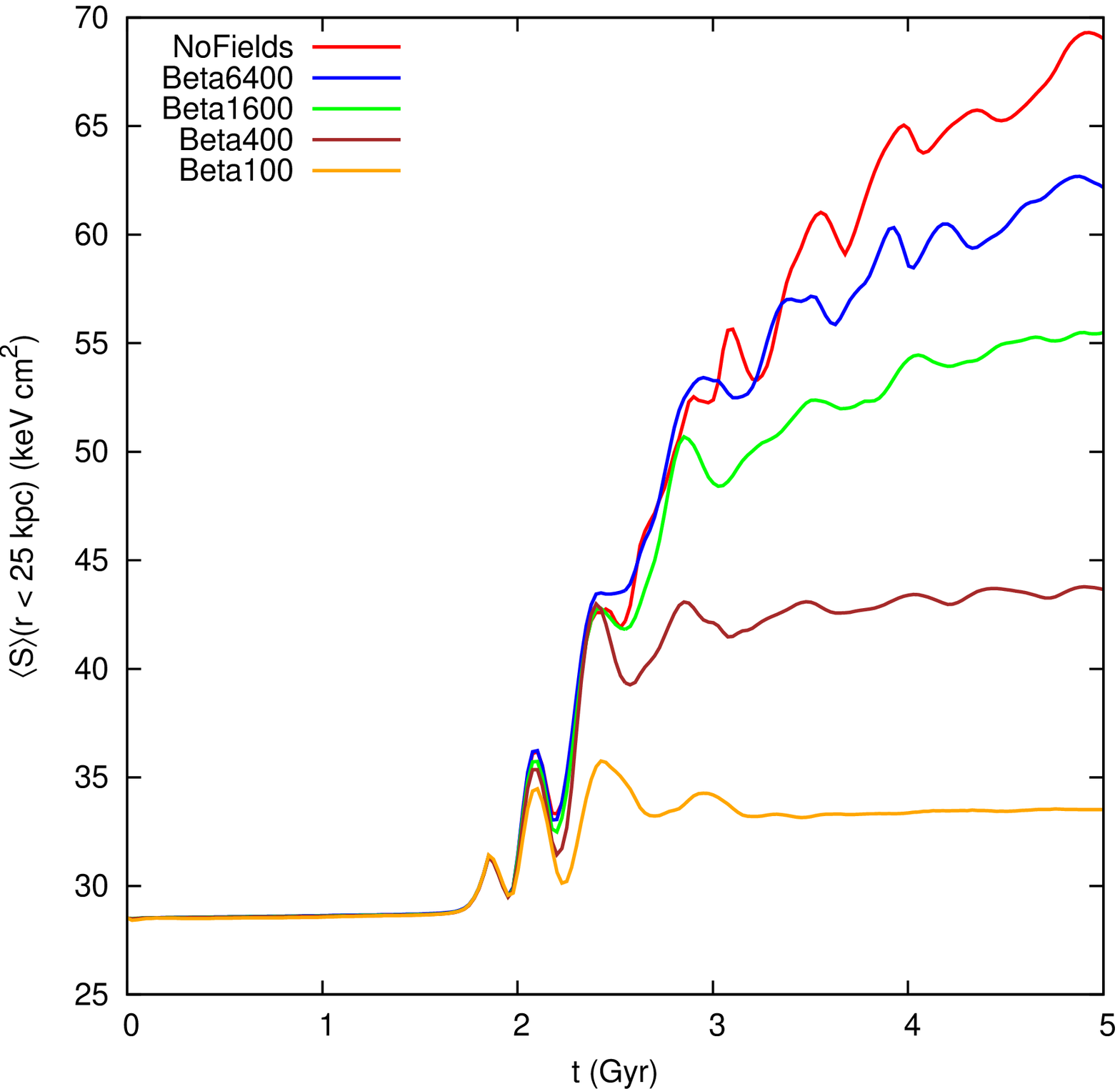}{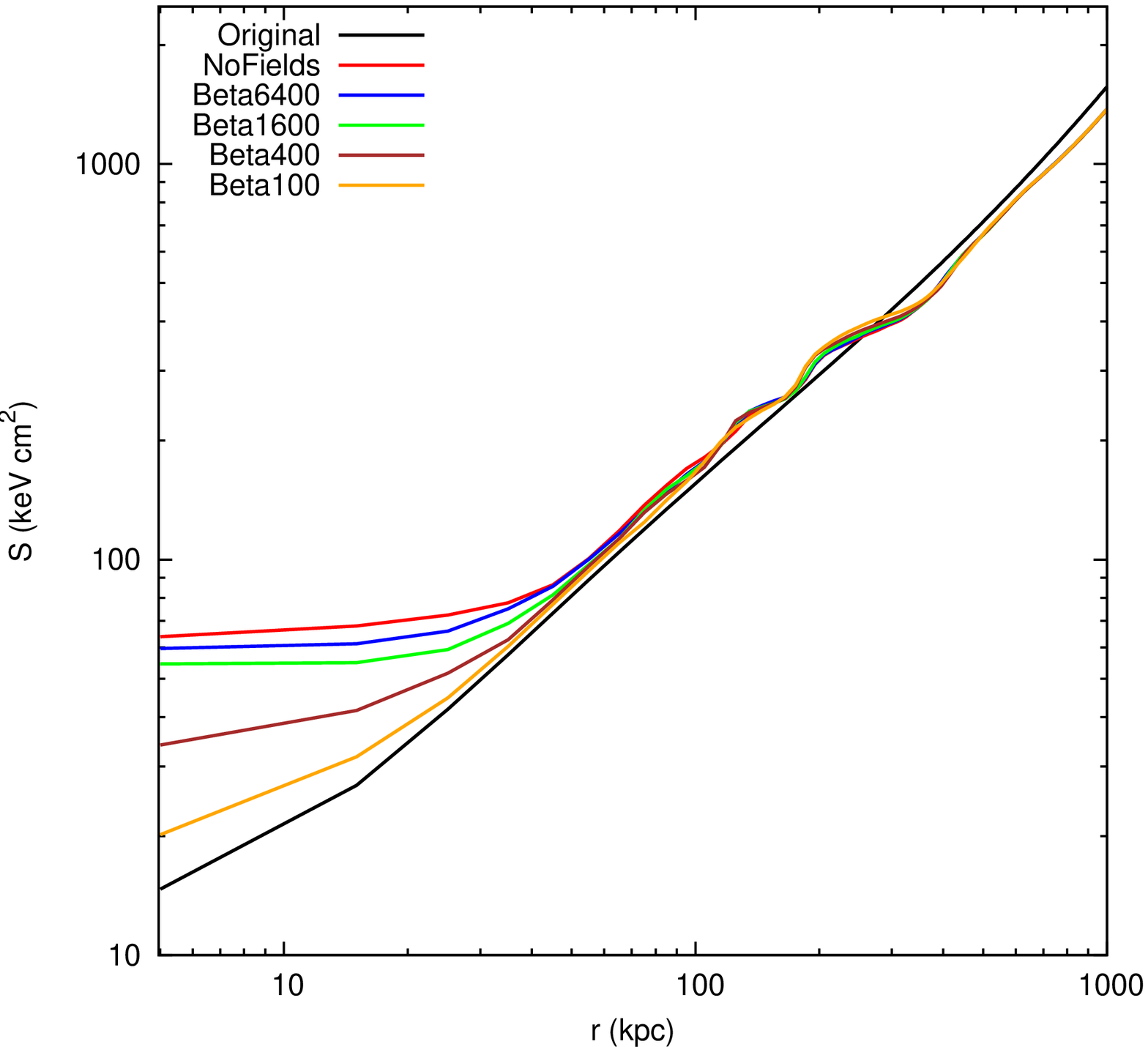}
\caption{The effect of magnetic fields of varying initial $\beta$ on the entropy of the gas. Left: Evolution of the average entropy within a radius of 25~kpc for the simulations with varying initial $\beta$, compared to the simulation with no magnetic fields. Right: Final entropy profiles (at $t$ = 5 Gyr) for the simulations with varying initial $\beta$, compared to the simulation with no magnetic fields.\label{fig:entropy}}
\end{center}
\end{figure*}

\begin{figure}
\begin{center}
\plotone{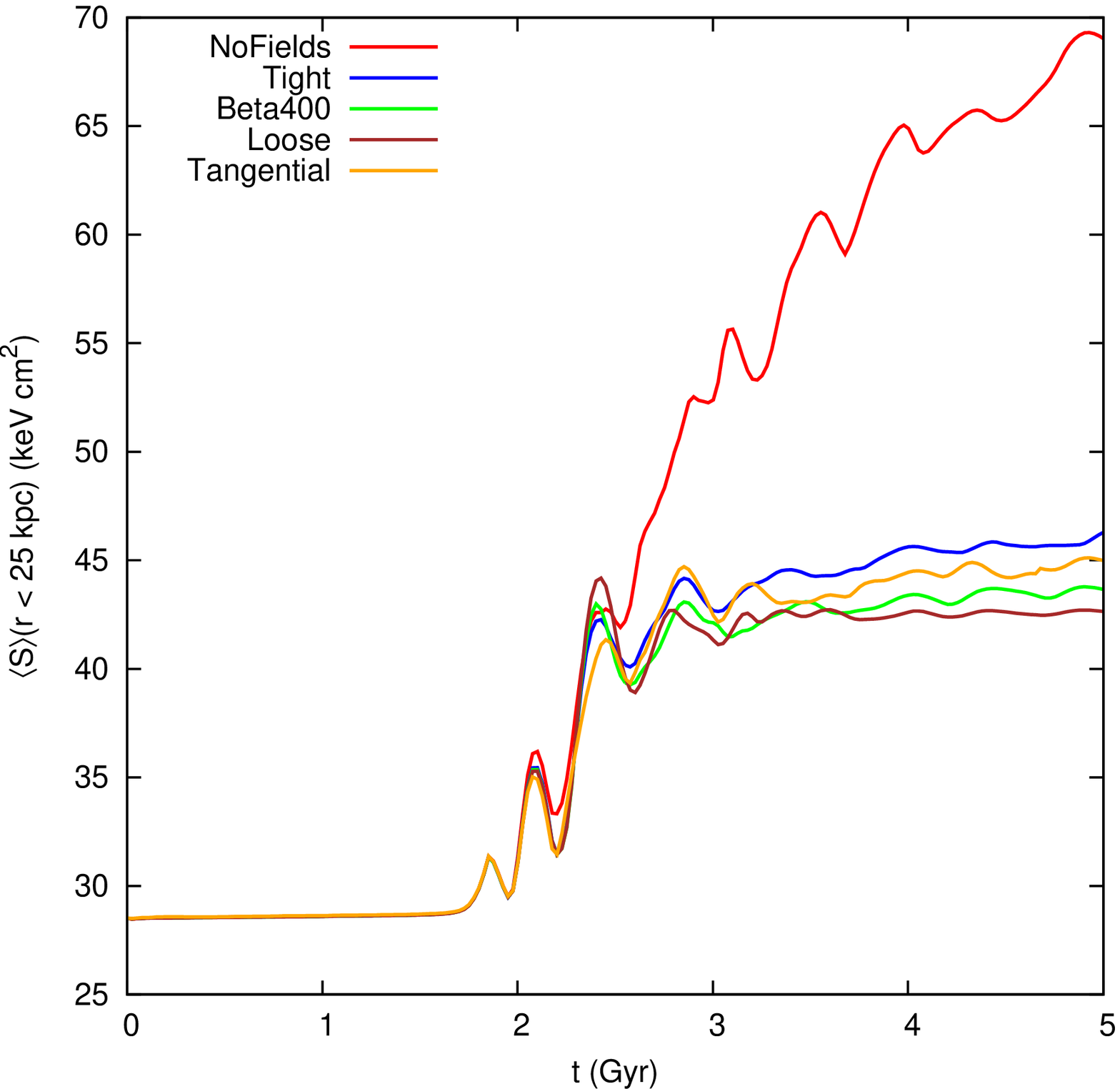}
\caption{Evolution of the average entropy within a radius of 25~kpc for the simulations with varying initial field configuration, compared to the simulation with no magnetic fields.\label{fig:entropy2}}
\end{center}
\end{figure}

\subsection{Cold Front Morphology in the Presence of Magnetic Fields\label{sec:withmag}}

We find that magnetic fields alter the gas dynamics of the sloshing cold fronts. We begin by examining the simulations that correspond to varying the initial plasma $\beta$. Figures \ref{fig:t2.5_temp} through \ref{fig:t4.5_temp} show temperature slices through the cluster center for the epochs $t$ = 2.5, 3.15, 3.75, and 4.5 Gyr after the beginning of the simulation for different initial $\beta$. Two effects are noticeable. The first is that as the magnetic field strength is increased, the cold fronts appear more smooth and regular, even though the instabilities are not suppressed completely. This is particularly true in the case of the fronts with the smallest radii ($r \sim$ 50~kpc) at later times. These fronts are smooth and well-defined in the simulations {\it Beta400} and {\it Beta100}, whereas they are more jagged and ill-defined in the simulations {\it Beta6400} and {\it Beta1600}. The fronts at large radii ($r \sim$ 150-200~kpc) in all simulations (with the exception of {\it Beta100}) are strongly affected by K-H instabilities (we will discuss the reason for this in Section \ref{sec:smooth_sharp_fronts}). The second significant effect is that as the magnetic field strength is increased, the cool gas in the very central part of the core ($r \simlt$ 50~kpc) is more resilient to the effects of sloshing. We will examine this effect more quantitatively in Section \ref{sec:mixing_and_heating}.

\begin{figure*}
\begin{center}
\plotone{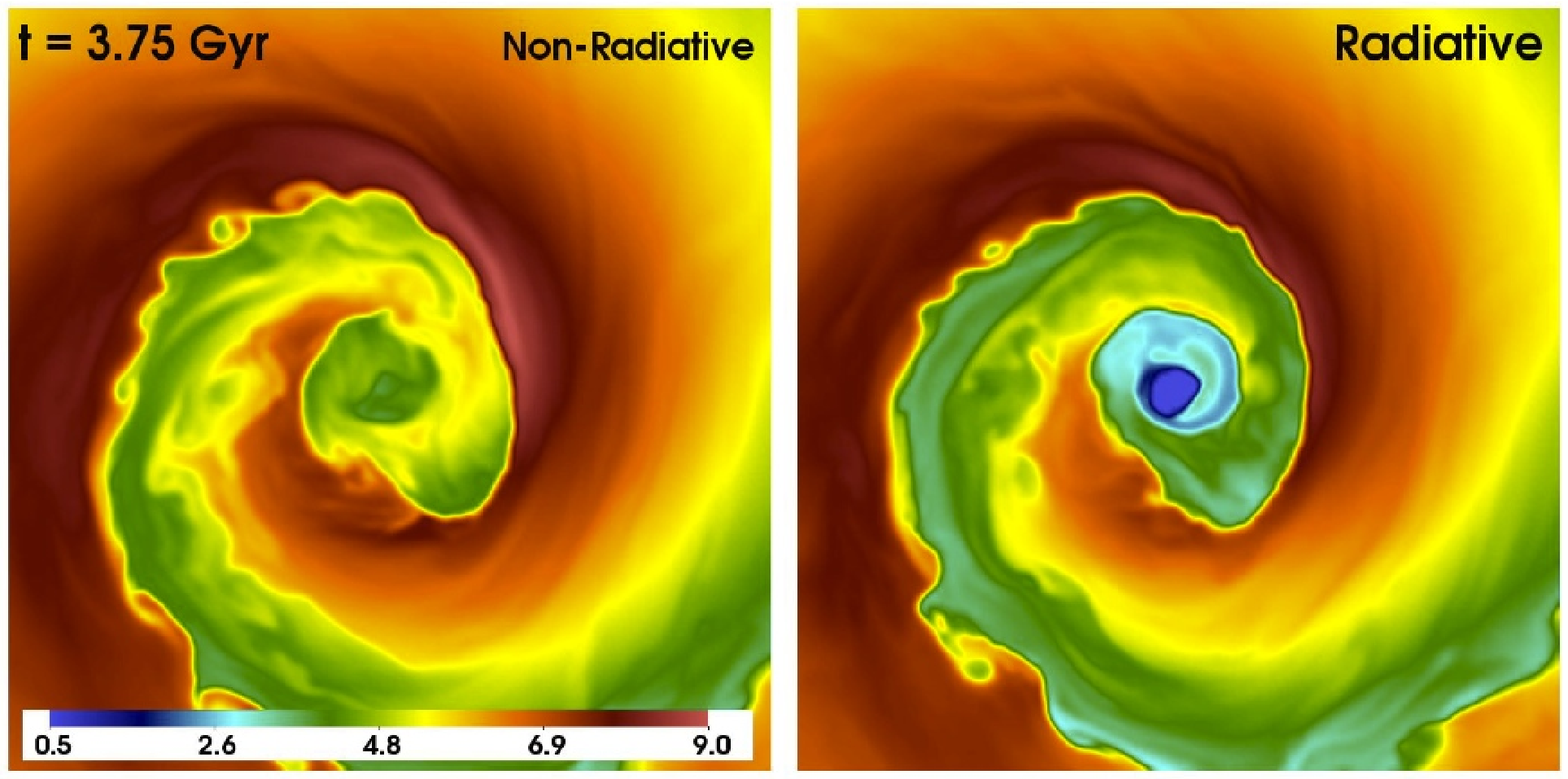}
\caption{Slices through the gas temperature for non-radiative and radiative simulations of gas sloshing at the epoch $t$ = 3.0~Gyr. The color scale shows the temperature in keV. Each panel is 500~kpc on a side.\label{fig:cooling1}}
\end{center}
\end{figure*}

\begin{figure}
\begin{center}
\plotone{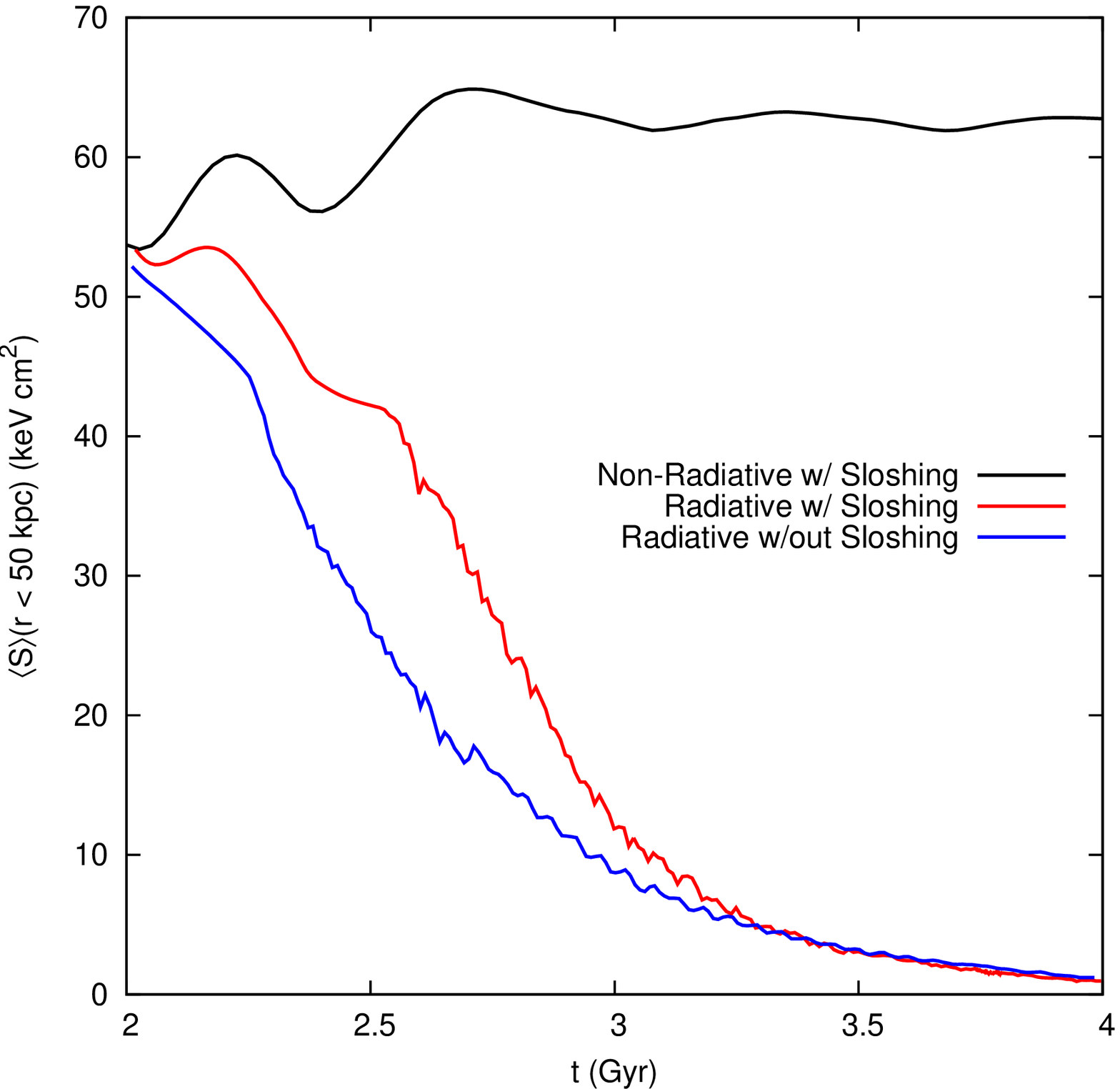}
\caption{Evolution of the average entropy within a radius of 50~kpc for simulations with and without radiative cooling.\label{fig:cooling2}}
\end{center}
\end{figure}

In contrast to the simulations where the initial magnetic field strength was varied, the effect of varying the initial magnetic field configuration is not as significant. Figure \ref{fig:t4.5_temp2} shows temperature slices through the cluster center for the epoch $t$ = 4.5 Gyr after the beginning of the simulation. The smoothness of the fronts is about the same for each of the simulations. The temperature contrasts across the fronts are also very similar, and the temperature of the core ($r \simlt$ 50~kpc) is approximately the same. In conformity with the results of the simulations with varying initial $\beta$, the cold fronts at large radii are most susceptible to the onset of the K-H instability.

One other particular aspect of the temperature maps deserves comment. Though in many cases, the front surfaces are smooth due to the large magnetic field amplification in their surrounding layers, the effect of the magnetic pressure can become dynamically important. This is particularly true in the case of the {\it Beta100} simulation, as in many places in this simulation, the magnetic pressure is a sizeable fraction of, or even comparable to, the gas thermal pressure (i.e., $\beta \rightarrow 1$). The result is fluctuations in density and temperature that anticorrelate with the fluctuations in magnetic pressure. An example of this is shown in Figure \ref{fig:fluctuations}. The sum of thermal and magnetic pressures is relatively smooth, whereas the gas thermal pressure itself shows small-scale variations. The resulting fluctuations in density and temperature are on the order of $\sim 10-30\%$ in this simulation.

\subsection{The Effect of Magnetic Fields on Mixing and Core Heating Due to Sloshing \label{sec:mixing_and_heating}}

Sloshing motions bring higher-entropy cluster gas at higher radii into contact with the low-entropy gas of the cool core, making it possible for these gases to mix and the average entropy per particle of the core to increase (ZMJ10). We have shown in the previous paper that in the absence of magnetic fields, the entropy of the core can increase significantly from its initial state. The left panel of Figure \ref{fig:entropy} shows the evolution in average entropy within a radius $r \leq 25$~kpc (well within the cooling radius for a relaxed cluster) for each of our simulations with varying $\beta$ and the {\it NoFields} simulation, the latter using the same hydrodynamical model as the invsicid simulations of ZMJ10. The center is taken to be the cluster potential minimum, and ``entropy'' is defined as $S \equiv k_BTn_e^{-2/3}$. Every simulation has an initial entropy per unit mass increase in the center associated with the passage of the subcluster and the removal of low-entropy gas from the cluster core. Following this initial increase, the evolution of the central entropy is strongly dependent of the details of the magnetic field. The right panel of Figure \ref{fig:entropy} shows the final entropy profiles of these simulations compared to the initial entropy profile. In the {\it NoFields} simulation, the cool core has been heated by sloshing and transformed to an isentropic core with a relatively high entropy. For the simulations with magnetic fields, as the initial magnetic field strength is increased, the final entropy profile changes less; even though sloshing brings hot and cold gases in contact, the magnetic field suppresses their mixing (as did isotropic viscosity in ZMJ10). 

In contrast, we do not find that varying the initial magnetic field spatial configuration changes the effectiveness of mixing. Figure \ref{fig:entropy2} shows the evolution in average entropy within $r \leq 25$~kpc for simulations with different initial field configurations (all with initial $\beta = 400$), as well as our control {\it NoFields} simulation. In each of the magnetized cases, the increase of entropy per unit mass of the gas in the cluster core is suppressed (in comparison to the {\it NoFields} simulation, but the degree of this suppression is essentially independent of the initial field configuration. This is expected, given the weak dependence of the final field configuration on the initial configuration that we noted in Section \ref{sec:bfield_amp}

In this analysis so far we have ignored the effects of radiative cooling. In ZMJ10, a series of sloshing simulations including radiative cooling were performed, to determine if the mixing of gases due to sloshing provided sufficient heat to the cluster core to offset a cooling catastrophe. It was found in that study that sloshing was able to heat the core for a short period, but that this heating was insufficient to completely offset radiative cooling. It was also found that if the ICM is significantly viscous, mixing of hot and cold gases would be suppressed, and radiative cooling would be offset by a nearly insignificant amount of heating. Since we find that the effect of the magnetic field is to similarly suppress mixing and the heating due to mixing, we expect that a cooling catastrophe would happen very quickly even in the presence of sloshing if the ICM is magnetized.

To confirm this, we have performed a simulation with radiative cooling included. The cooling function is calculated using the MEKAL model \citep{MKL95}, assuming the gas has a uniform metallicity $Z = 0.3Z_{\odot}$. Otherwise, the simulation is identical to the {\it Beta400} simulation. We have followed the strategy of ZMJ10 in beginning the simulation shortly after core passage (corresponding to the epoch $t$ = 2.0~Gyr in the non-radiative simulations), to avoid the compression of the cluster core that occurs during this period, which would lead to significant overcooling and hasten a cooling catastrophe. Our aim is to determine the effect of sloshing on the cooling cluster core in isolation. To compare with this simulation, we have also ran a simulation of an isolated, magnetized galaxy cluster with radiative cooling. We follow the simulations for 2~Gyr.

Figure \ref{fig:cooling1} shows temperature slices at the epoch $t$ = 3.75~Gyr for the radiative and non-radiative sloshing simulations. The most obvious difference between the two simulations is the presence of the large temperature drop ($T < 1$~keV) in the center that has occurred as a result of the unabated cooling of the cluster core gas. The continued presence of low-entropy gas has also resulted in cold fronts with a higher temperature contrast when compared to the non-radiative case. Figure \ref{fig:cooling2} shows the evolution of the average entropy with a radius of $r = 50$~kpc for the radiative and non-radiative sloshing simulations, compared to the radiative simulation of a cluster in isolation. The entropy of the core in the radiative simulation drops very quickly, and is only modestly slowed down by the heat contribution from sloshing when compared to the case when there is no sloshing at all. This shows that the presence of the magnetic field suppresses mixing of hot and cold gases to such a degree that the effect of sloshing on the cooling rate of the core is very modest.  In line with the results from ZMJ10, we find that the heat delivered from sloshing alone is unable to stave off a cooling catastrophe, especially in the case of a magnetized ICM. However, thermal conduction has not been considered in these simulations. Since sloshing brings hot and cold gases into close contact, it may yet result in significant heating of the cluster core due to thermal conduction. An investigation of core gas sloshing including the effects of anisotropic thermal conduction is the subject of a future paper (ZuHone et al. 2011, in preparation). 

\subsection{Mock Observations\label{sec:mock_obs}}

The observational signatures of sloshing cold fronts are the spiral-shaped bright edges in the X-ray emission in the cluster cores. To compare our simulations more directly with these observations, we have constructed mock X-ray surface brightness observations from our simulation data. Each cell in the AMR grid has a photon emission (in photons~s$^{-1}$~cm$^{-3}$) given by 
\begin{equation}
\epsilon_\gamma = n_en_H\Lambda_\gamma(T,Z)
\end{equation}
where $n_e$ and $n_H$ are the electron and hydrogen densities, respectively, and $\Lambda_\gamma(T,Z)$ is the emissivity which depends on temperature and metallicity, which are assumed constant over one FLASH cell size. The emissivity is calculated using the MEKAL model \citep{MKL95}, under the assumption that the cluster is situated at redshift $z = 0.06$ and a constant metallicity of $Z = 0.3Z_{\odot}$, an assumption adequate for our qualitative comparisons. 
Using this relation, the photon luminosity of each sky pixel in photons~s$^{-1}$ integrated over the chosen energy range (in this case, the {\it Chandra} band of 0.5-7.0~keV in the observer's frame) and projected along the line of sight is given by
\begin{equation}
L_\gamma= \int_V{\epsilon_\gamma}dV' \approx \sum_i{\Lambda_{\gamma,i}}n_{e,i}n_{H,i}\Delta{V_i}
\end{equation}
where the subscripts $i$ refer to the quantities in each AMR cell. The resulting surface brightness map is uniformly gridded to the resolution of the smallest cell in the simulation, $\Delta{x} \sim 2$~kpc. 

\begin{figure}
\begin{center}
\plotone{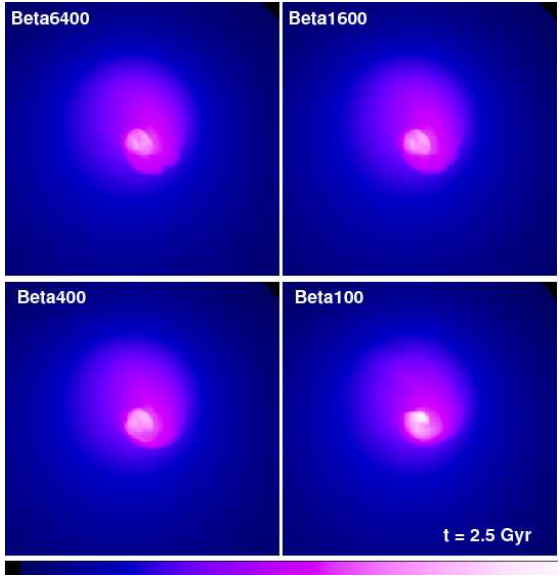}
\caption{Projected X-ray brightness (in the 0.5-7.0 keV band) for the simulations with varying $\beta$ at the epoch $t$ = 2.5~Gyr. The brightness scale is square root and the same for each panel. Each panel is 500~kpc on a side.\label{fig:t2.5_xray}}
\end{center}
\end{figure}

\begin{figure}
\begin{center}
\plotone{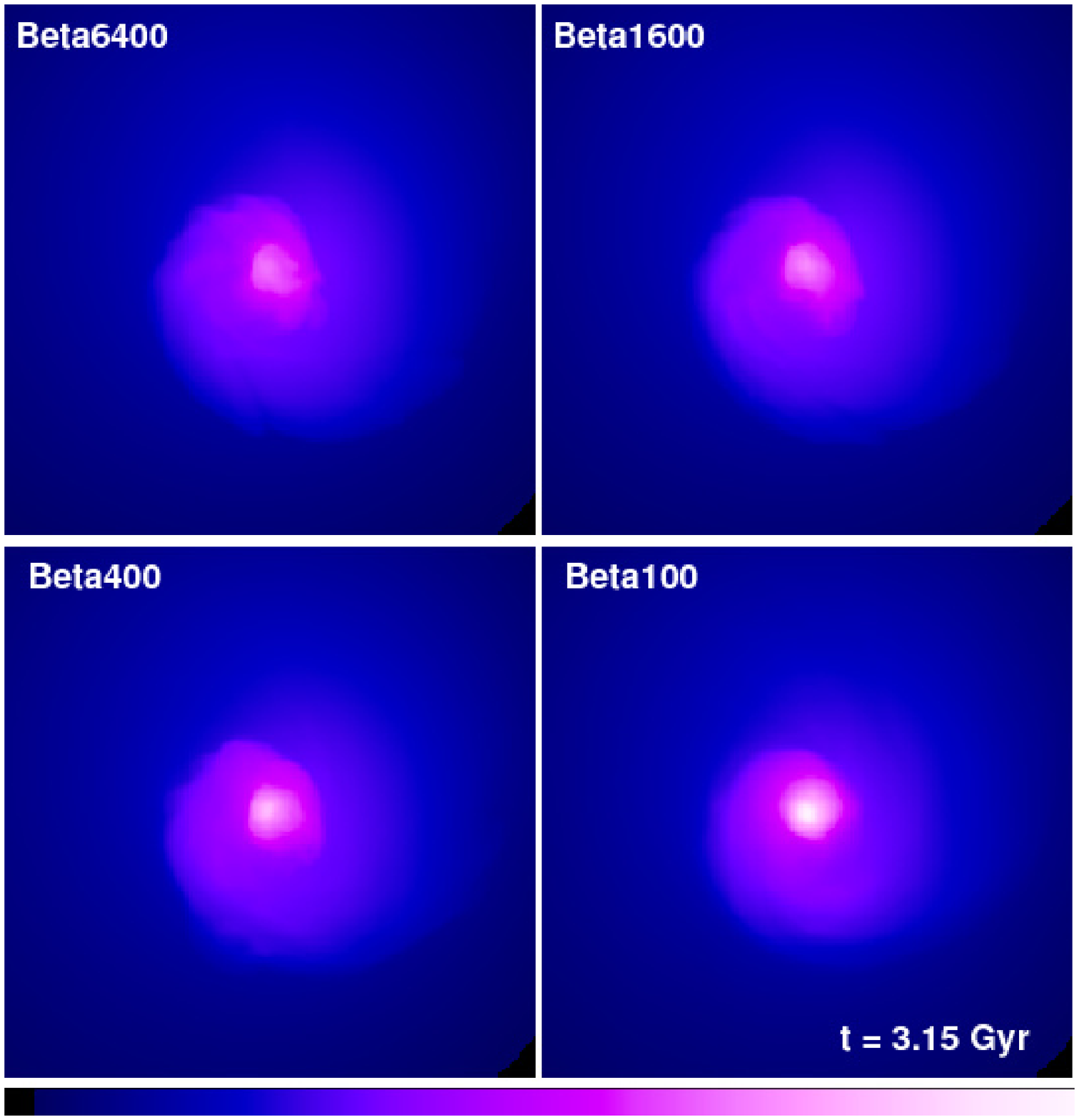}
\caption{Projected X-ray brightness (in the 0.5-7.0 keV band) for the simulations with varying $\beta$ at the epoch $t$ = 3.15~Gyr. The brightness scale is square root and the same for each panel. Each panel is 500~kpc on a side.\label{fig:t3.15_xray}}
\end{center}
\end{figure}

\begin{figure}
\begin{center}
\plotone{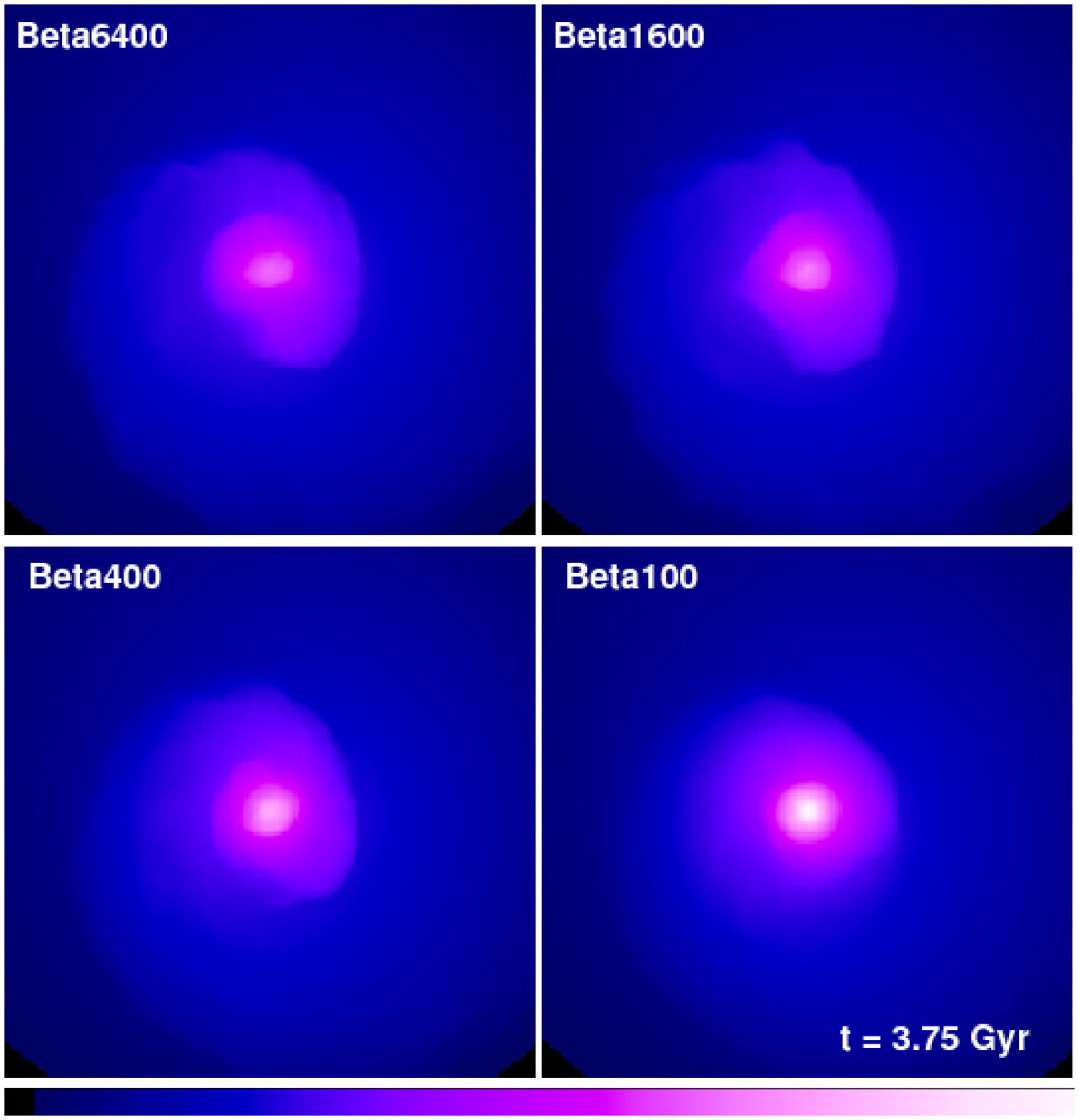}
\caption{Projected X-ray brightness (in the 0.5-7.0 keV band) for the simulations with varying $\beta$ at the epoch $t$ = 3.75~Gyr. The brightness scale is square root and the same for each panel. Each panel is 500~kpc on a side.\label{fig:t3.75_xray}}
\end{center}
\end{figure}

\begin{figure}
\begin{center}
\plotone{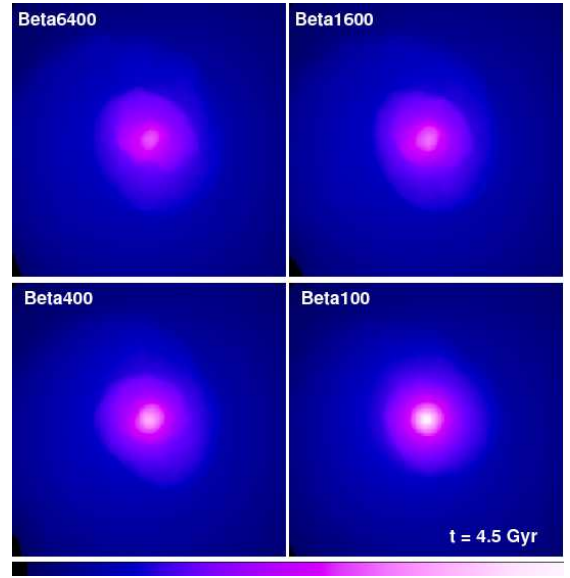}
\caption{Projected X-ray brightness (in the 0.5-7.0 keV band) for the simulations with varying $\beta$ at the epoch $t$ = 4.5~Gyr. The brightness scale is square root and the same for each panel. Each panel is 500~kpc on a side.\label{fig:t4.5_xray}}
\end{center}
\end{figure}

Figures \ref{fig:t2.5_xray} through \ref{fig:t4.5_xray} show the resulting X-ray surface brightness maps for the simulations with varying $\beta$ for the epochs $t$ = 2.5, 3.15, 3.75, and 4.5~Gyr after the beginning of the simulation. There is a trend of increasing smoothness of cold fronts with increasing magnetic field. Additionally, the simulations with higher magnetic field have a brighter central core, in keeping with the result from Section \ref{sec:mixing_and_heating} that the cool core of the cluster is maintained in these simulations. These results are similar to the results that were obtained in ZMJ10 when viscosity was implemented. A more quantitative comparison with observations will be given in a future paper.

\section{Discussion\label{sec:disc}}

\subsection{A Comparison of Amplified Field Strengths with Earlier Estimates\label{sec:amp_comparisons}}

Under the assumptions of ideal MHD, the magnetic field lines are ``frozen'' into the flow, a condition that can be expressed by the equation \citep{cha61}:
\begin{equation}
\frac{\partial{\bf B}}{\partial{t}} + \nabla \cdot ({\bf v}{\bf B} - {\bf B}{\bf v}) = 0
\end{equation}
which, combined with the continuity equation, gives
\begin{equation}
\frac{d}{dt}\left(\frac{\bf B}{\rho}\right) = \left(\frac{\bf B}{\rho} \cdot \nabla\right){\bf v},
\end{equation}
which implies that along shear flows the magnetic field will be stretched and amplified. \citet{kes10} derived an analytic estimate for the shear amplification of the magnetic field along cold fronts. From Equation 11 from \citet{kes10}, we have (under the assumption of compressibility):
\begin{equation}
\frac{B_\phi}{B_{r'}} \sim t\partial_{r'}v \sim 10\M_iT_4^{1/2}\left(\frac{\Delta}{\rm 10~kpc}\right)^{-1}\left(\frac{t}{10^8~{\rm yr}}\right)
\end{equation}
where $\M_i$ is the Mach number of the flow inside the front, $T_4 \equiv T$/(4~keV) is the temperature inside the front, $\Delta$ is the thickness of the shear layer in kpc, and $t$ is the development time of the magnetization layer. $B_\phi$ is the amplified field strength just under and parallel to the front, and $B_{r'}$ is the initial magnetic field perpendicular to the front surface. To compare with the amplification of $\beta$ seen in our simulations, we first note that for an initially random, tangled field the magnetic energy in the component perpendicular to the front surface is roughly one-third of the magnetic energy in the total field, $B_{r'}^2/8\pi \sim (1/3)B^2/8\pi$. For the final amplified field, we expect that $B_\phi^2/8\pi \sim B^2/8\pi$. Setting $B_i = \sqrt{3}B_{r'}$ and $B_f = B_\phi$, assuming $B_f^2 \gg B_i^2$, and rearranging in terms of $\beta$, we find:
\begin{equation}
\frac{\beta_f}{\beta_i} \sim \frac{0.03}{\M_i^2T_4}\left(\frac{\Delta}{\rm 10~kpc}\right)^2\left(\frac{t}{10^8~{\rm yr}}\right)^{-2}
\end{equation}
where $\beta_i$ and $\beta_f$ are the initial and final plasma $\beta$. Taking representative numbers from our simulations, we have $\M_i \sim 0.3-0.5$ for the Mach number and $T \sim 4$~keV for the temperature just inside the cold front. In our simulations the thickness of the shear layer is $\Delta \sim 5-10$~kpc, and the amplified magnetic layers typically take a few $\times 10^8$ yr to develop. Under these conditions, the decrease in $\beta$ should be $\beta_f/\beta_i \sim 0.1 - 0.01$, correspondng to a magnetic field energy amplification $B_f^2/B_i^2 \sim 10-100$, which is agreement with what we find in our simulations, corroborating the results of \citet{kes10}. 

\subsection{Smooth and Sharp Cold Fronts with Magnetic Fields\label{sec:smooth_sharp_fronts}}

Our simulations demonstrate that for magnetic field strengths compatible with those inferred from observations, sloshing will result in shear amplification and stretching of the magnetic field lines along the cold fronts. The stretched and amplified fields will suppress instabilities of the fronts and help preserve the fronts' smooth shapes. The efficacy of this is somewhat dependent on the initial strength of the magnetic field, as stronger initial fields are amplified more quickly to strengths that can suppress the instabilities.

In addition to suppressing these instabilities, the magnetic fields have another related effect which results in the persistence of smooth and high-contrast cold fronts. By suppressing the mixing of high and low-entropy gases, the magnetic fields ensure that the densest gas in the cluster remains cold compared to the surrounding gas. \citet{ghi10} pointed out a correlation between clusters with sloshing cold fronts and clusters with steep entropy gradients in the core. This was predicted in AM06; if the entropy gradient in the cluster core is not significant, the entropy contrast is not sufficient for the cool gas pushed out of the disturbed cluster potential minimum to flow back and for sloshing to begin. Similarly, sloshing will persist in generating cold fronts with higher contrast if the steep entropy gradient is maintained during the sloshing period. Since sloshing mixes high and low-entropy gas, possibly eliminating this gradient (ZMJ10), mixing should be suppressed in order to maintain the sloshing. Our simulations indicate that the stronger the magnetic field is, the longer the original steep entropy gradient is maintained. 

In V01/V02, the lack of evidence for the growth of the Kelvin-Helmholtz instability on the surface of a prominent cold front in the merging galaxy cluster A3667 was used to argue for the existence of a magnetic field of $B \sim$ 10~$\mu$G parallel to the front. This was on the basis of a simple stability analysis for tangential perturbations on a shearing surface in the presence of a magnetic field. It is instructive to see how the same analysis fares when applied to our simulated cold fronts. In particular, can the magnitude of the actual magnetic field strength in our simulation be predicted? For this purpose we have chosen a few fronts in our simulations and have carried out the same rough calculation, made more straightforward by the fact that we have direct access to the relevant quantities (density, velocity, etc.). 

\begin{table*}[thdp]
\caption{Cold Front Stabillity Analysis\label{tab:fronts}}
\begin{center}
\begin{tabular}{ccccc}
\hline
\hline
Front & $\Delta{v}$ (km/s) & $\rho_c/\rho_h$ ($10^{-26}$~g~cm$^{-3}$) & $B_{\rm pred}$ ($\mu$G) \\
\hline
1 & 250 & 4.0/1.5 & 13.1 \\ 
2 & 150 & 6.5/3.5 & 8.0 \\
\hline
\hline
Front & \multicolumn{4}{c}{$B_{\rm sim}$ ($\mu$G)} \\
\cline{2-5}
& ${\rm Beta100}$ & ${\rm Beta400}$ & ${\rm Beta1600}$ & ${\rm Beta6400}$  \\
\hline
1 & 19.4 & 11.2 & 5.0 & 1.1 \\
2 & 15.9 & 7.1 & 5.0 & 3.2 \\
\hline 
\end{tabular}
\end{center}
\end{table*}

The dispersion equation for small tangential-discontinuity perturbations in a perfectly conducting, incompressible plasma can be written as \citep[][in Gaussian units]{lan60}
\begin{equation}
\rho_h(\omega - k\Delta{v})^2+ \rho_c\omega^2 = k^2\left(\frac{B^2_h}{4\pi} + \frac{B^2_c}{4\pi}\right)
\end{equation} 
where $B_h$ and $B_c$ are the magnetic field strengths in the hot and cold gases, respectively, $\rho_h$ and $\rho_c$ are the gas densities in the hot and cold gases, $v$ is the shear velocity difference across the front, and $\omega$ and $k$ are the perturbation frequency and wavenumber. The discontinuity is stable if \citep[V01, V02,][]{kes10}
\begin{equation}
\frac{B^2_h}{8\pi} + \frac{B^2_c}{8\pi} > \frac{1}{2}\frac{\rho_h\rho_c}{\rho_h+\rho_c}(\Delta{v})^2
\end{equation}
The treatment of the gas as incompressible, which simplifies the analysis, is justified by the fact that the Mach numbers of the gas flows in these regions are relatively low ($M \simlt 0.5$), and the the growing modes of the K-H instability in the cool gas have low phase speed.

\begin{figure*}
\begin{center}
\plottwo{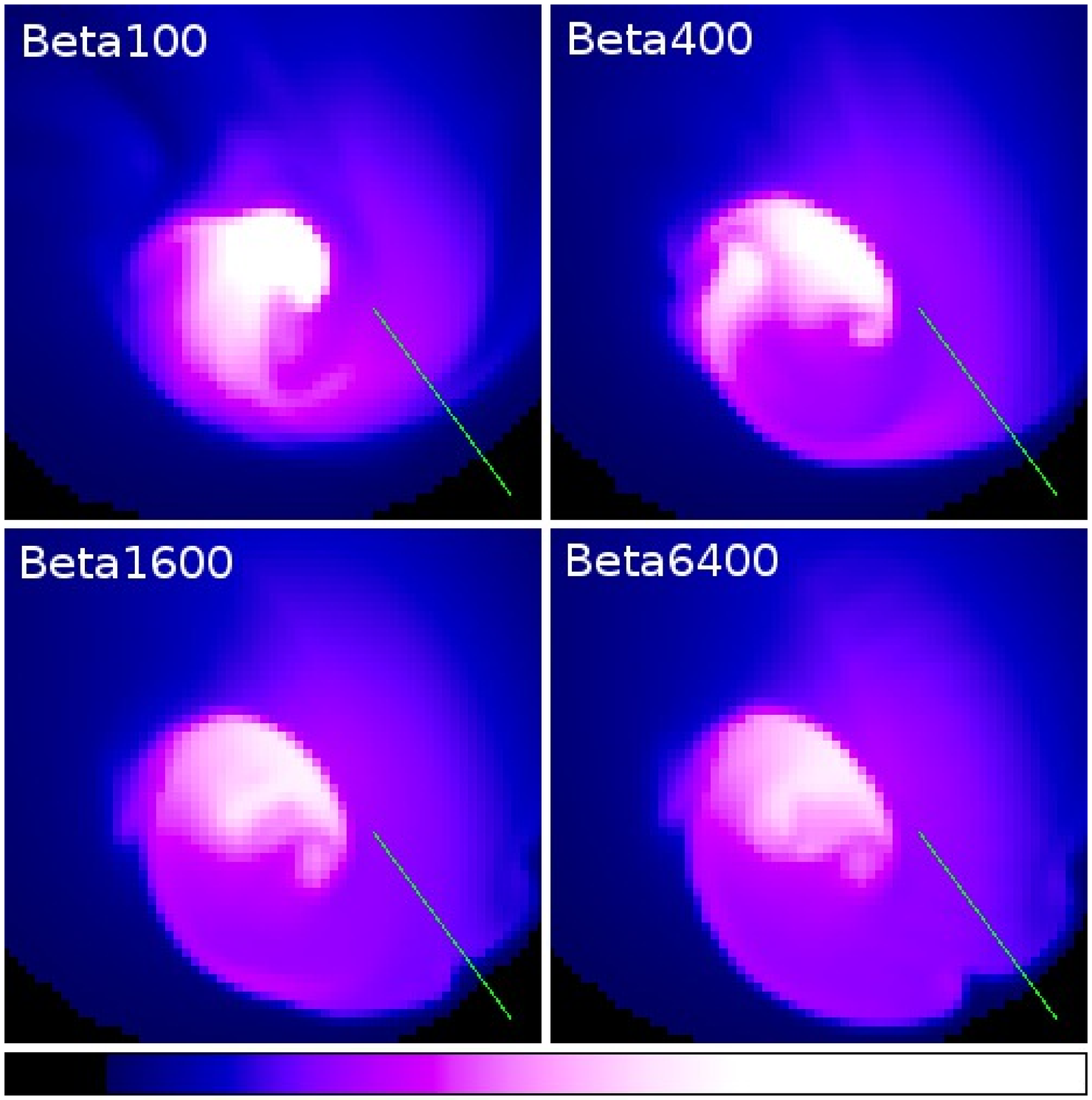}{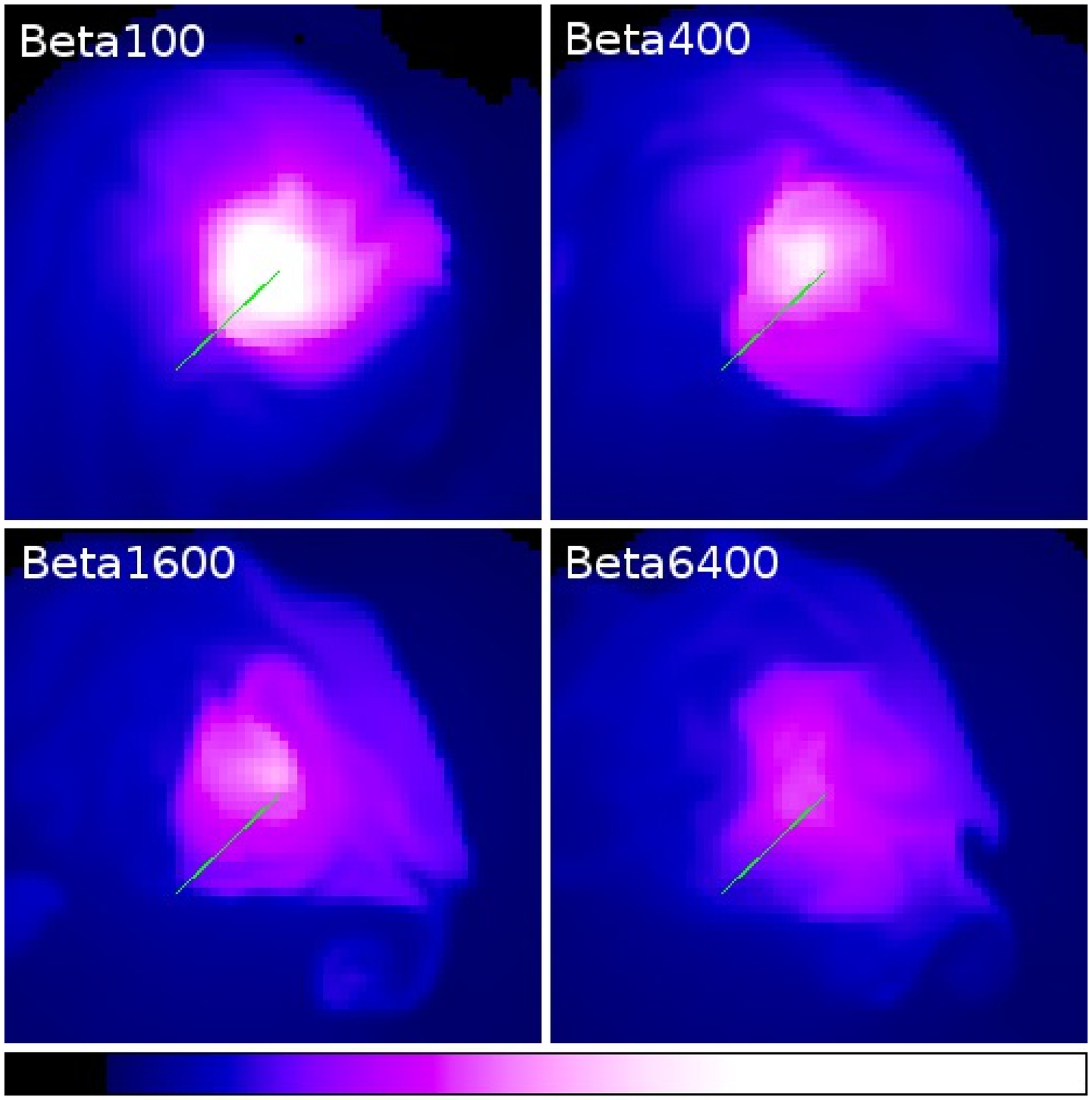}
\caption{Fronts chosen for stability analysis. Plots are of gas density, where the scale is the same for all panels. The fronts chosen are marked with a green line. Left set of panels: Front 1, for the simulations with varying $\beta$ at the epoch $t$ = 2.5~Gyr. Right set of panels: Front 2, for the epoch $t$ = 3.15~Gyr. Each panel is 150 kpc on a side. The color scale is different from that in Figures \ref{fig:t2.5_xray} through \ref{fig:t4.5_xray}.\label{fig:fronts}}
\end{center}
\end{figure*}

We examine two fronts from our simulations with varying initial $\beta$, one that appears at the epoch $t$ = 2.5~Gyr and another that appears at the epoch $t$ = 3.15~Gyr. These fronts have been singled out due to their smooth appearance in the simulations with higher magnetic field, but the evidence of instability when the magnetic field strength is lower. The two fronts and the front cross-sections along which we examine the gas quantities are shown in Figure \ref{fig:fronts}. In the simulations {\it Beta100} and {\it Beta400}, the chosen fronts are smooth and sharp, but in the simulations {\it Beta1600} and {\it Beta6400} the fronts are visibly disturbed by instabilities.  

For both fronts, for simplicity we have assumed that the densities across each front and the velocity shear are the same for all four simulations with different initial $\beta$, which is accurate to approximately 10\% and is sufficient for our current purpose. Table \ref{tab:fronts} shows the densities, shear velocities, and predicted minimum magnetic field strengths for the two fronts if they are stable, and the actual total magnetic field strengths in the four simulations {\it Beta100,Beta400,Beta1600} and {\it Beta6400}. 

In the {\it Beta100} simulation, the magnetic fields at the front surfaces are stronger than the minimum value required for stability, consistent with the stability of the fronts. In the case of the {\it Beta400} simulation, the fronts are stable, while the actual magnetic field strengths are very close to the values necessary to stabilize the front. In the {\it Beta1600} and {\it Beta6400} simulations, the fronts are not stable, and the corresponding magnetic field strengths are far less than the values required for stability. 

Strictly speaking, this analysis is only valid in the incompressible limit for a plane-parallel surface \citep[see][for an alternative hypothesis for front stability based on the front curvature]{chu04}. Given the approximations of the above qualitative stability estimate, the predicted field estimates based on this stability analysis agree quite well with the actual field strengths that are capable of stabilizing the fronts.

\subsection{Amplification of Magnetic Fields by Sloshing: Implications\label{sec:amp_implications}}

\begin{figure*}
\begin{center}
\plotone{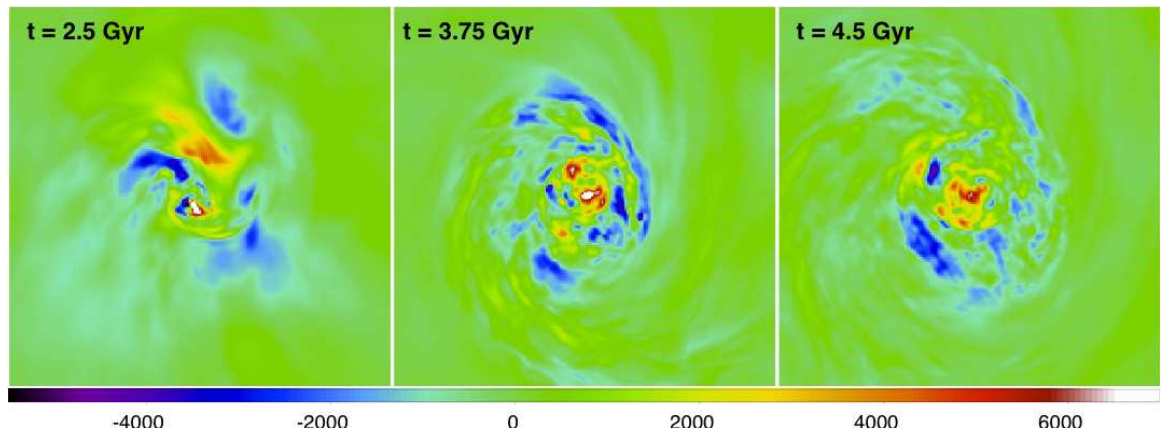}
\caption{Simulated rotation measure maps for selected epochs in the {\it Beta400} simulation. Each panel is 500~kpc on a side.\label{fig:RM_maps} Color scale shows rotation measure in rad/m$^2$.}
\end{center}
\end{figure*}

Magnetic fields amplified by sloshing motions may have other observable effects. One such effect that has been suggested by previous works is that of radio mini-halos. Mini-halos are regions of diffuse synchrotron emission found in the cooling core regions of some relaxed clusters. A possible connection to mini-halos and sloshing was suggested by \citet{maz08}, who discovered a correlation between the radio minihalo emission and the regions bounded by the sloshing cold fronts in two galaxy clusters. Other clusters with evidence of sloshing cold fronts also sport minihalos, including RXC J1504.1-0248 \citep{gia11} and Perseus (ZuHone et al. 2011, in preparation).

The existence of radio mini-halos necessitates a source for the relativistic electrons. Two models have been proposed: the hadronic model \citep{den80,ves82,bla99,dol00,pfr04} and the reacceleration model \citep{cas05,cas07,bru11}. In the hadronic model, relativistic electrons are produced as byproducts of hadronic interactions of cosmic ray protons with thermal protons in the ICM, the resulting electrons producing synchrotron radio emission. The coincidence between the radio emission and the cold fronts in the hadronic model scenario would be due to the shear-amplified magnetic fields that are produced as a result of the sloshing, as suggested by \citet{kes10} and seen in our simulations. Alternatively, MHD turbulence driven by the sloshing motions may accelerate relativistic electrons via damping of magnetosonic waves \citep{eil79,bru07}. Numerical experiments to explore the relationship between gas sloshing in relaxed clusters and radio mini-halos is the subject of a forthcoming paper (ZuHone et al. 2011, in preparation). 

Since the sloshing motions in our simulations create strong magnetic fields ordered on large scales, it may be possible to detect these fields via Faraday rotation measurements. A field line directed along our line of sight will produce a rotation measure of the polarized radio emission given by 
\begin{equation}
{\rm RM[rad~m}^{-2}] = 812\int_0^L{n_e}{B_\parallel}{\rm d}l
\end{equation}
where $n_e$ is the electron number density in cm$^{-3}$, $B_\parallel$ is the parallel magnetic field strength in $\mu$G, and $L$ is the length of the source in kpc along the line of sight. If the field lines in the amplified layers are partially directed along the line of sight, it should be possible to detect them in rotation measure observations of background radio galaxies. In Figure \ref{fig:RM_maps}, we give examples of simulated RM maps from the {\it Beta400} simulation at three different epochs. In the early stages of sloshing (left and center panels), long, coherent structures in the RM map with large values are coincident with the places where the cold fronts can be seen in X-rays. At later times, the layers are not as prominent, as the field amplification is weaker and the field within the cold fronts is more tangled (right panel). The values of the rotation measure in these maps are comparable to those seen in cool-core clusters, such as Perseus \citep{tay06} and Hydra A \citep{tay93}. The spatial coincidence of a background radio galaxy with a cold front may be a prime opportunity to find some observational confirmation of a strong magnetic layer along a cold front. Alternatively, the CMB itself may be used as the source of polarized photons \citep{ohn03}, possibly allowing estimates of the magnetic field strength to be made over a large spatial area of the cluster. Due to the very specific alignment of magnetic field layers with sloshing cold fronts, and the results of this work suggest that clusters with observed cold fronts would be good candidates for such a study. 

Caution should be taken, however, when interpreting these results. The accuracy of magnetic field estimates from the RM method depends on the statistical independence of the fluctuations in magnetic field and the thermal electron density. Anticorrelated fluctuations between the electron density and the magnetic field strength will result in an underestimate of the magnetic field in these regions, since lower-density fluctuations will be weighed less \citep{bec03}. This could be a possible concern for detecting the magnetized layers seen in our simulations, as Figure \ref{fig:fluctuations} shows that they tend to be associated with fluctuations of underdense gas. The fluctuations that we see are on the order of $\sim$10\%, so this should not be a strong effect. In any case, the magnetic fields in these layers will likely be the strongest along the line of sight, so they should still be easily detectable despite this possible systematic effect. 

\section{Summary\label{sec:summary}}

We used high-resolution magnetohydrodynamic simulations of gas sloshing in the cluster cool cores initiated by the infall of subclusters. We study the effect of such sloshing on the magnetic field in the intracluster medium and the effect of this field on the cold fronts seen in X-ray observations of relaxed galaxy clusters. We explored a range of initial conditions for the magnetic field, including varying its initial strength and spatial configuration.

Our results show that as a result of the shear flows accompanying the sloshing motions, the magnetic field energy is increased significantly along the cold front surfaces. The degree of amplification of the magnetic field strength along these surfaces is up to an order of magnitude, resulting in an overall energy of the field $B^2/8\pi$ that is amplified by up to an order of magnitude or two from its average value prior to the onset of sloshing. In particular, in the layers along the cold fronts, fields with initial strengths of 0.1-1~$\mu$G may be amplified to tens of $\mu$G. The final strength of the magnetic field is dependent on the initial field strengths, but weakly dependent on the initial spatial configuration of the field. Our results for the field strengths along the fronts are in agreement with previous analytic estimates \citep{kes10}. The sloshing motions result in a magnetic field that is tangled on small scales within the cluster cool core, and ordered structures along the fronts, regardless of the initial field configuration. Importantly, we find that regardless of the initial field strength or configuration, the final field strength averaged over the sloshing region is very similar, implying the field strength does not increase without limit due to amplification by sloshing, but saturates.

If the increase of the magnetic field strength is high enough, our simulations show that perturbations that are able to grow due to the Kelvin-Helmholtz instability and disrupt the front surfaces are suppressed. The degree of suppression of these perturbations is highly dependent on the magnetic field strength of the cluster. Because the degree of amplification of the magnetic field is similar across the simulations, the simulations with the highest initial magnetic field strength are most effective at maintaining smooth front surfaces. There is very little dependence of the effectiveness of the field to preserve the fronts on the initial spatial configuration of the magnetic field, because the final field does not remember much of its initial configuration. The cold fronts at high radii are most susceptible to the onset of the K-H instability, mainly due to the weaker magnetic fields at those radii. 

The ordered magnetic fields also suppress mixing in the intracluster medium. Sloshing in a cool-core cluster brings hot and cold gas in close contact. In the absence of these magnetic fields, these gases mix because of the development of instabilities, resulting in a net increase in entropy per gas particle in the core (ZMJ10). We find that as the average magnetic field strength in the cluster is increased, the mixing of gases is inhibited, and the sloshing-induced heating of the core is hindered. The simulations with the strongest magnetic fields result in very little change to the radial entropy profile of the cluster. When cooling is included, we find that the magnetic fields suppress mixing to such a degree that the heat contributed to the cluster core from sloshing is negligible. 

The interplay of sloshing motions and magnetic fields in relaxed clusters may have important consequences for future simulations and observations. The strong magnetic fields, ordered on large scales, produced by sloshing could potentially be detected in rotation measure maps, if a radio galaxy happens to be located behind a front in the plane of the sky.  Several examples exist of radio minihalos associated with X-ray cold fronts in relaxed clusters. Our simulations show that within the sloshing region, magnetic fields are amplified, which could be partially responsible for the correspondence between the radio emission and the X-ray fronts \citep{kes10}. The sloshing motions are also likely to generate turbulence, and it is possible that this turbulence reaccelerates an existing population of relativistic electrons, which will be addressed in a future paper (ZuHone et al. 2011, in preparation).

\acknowledgments
We would like to thank the anonymous referee for helpful and constructive comments and suggestions. JAZ thanks Ian Parrish, Bill Forman, Eric Hallman, and Mateusz Ruszkowski for useful discussions and advice. Calculations were performed using the computational resources of the Smithsonian Insitution's Hendron Data Center, Argonne National Laboratory, and the National Institute for Computational Sciences at the University of Tennessee. Analysis of the simulation data was carried out using the AMR analysis and visualization toolset yt \citep{tur11}, which is available for download at \url{http://yt.enzotools.org}, and for which Matthew Turk provided considerable help toward getting it working for the analysis for this paper. JAZ is supported by the NASA Postdoctoral Program. The software used in this work was in part developed by the DOE-supported ASC / Alliances Center for Astrophysical Thermonuclear Flashes at the University of Chicago.

\appendix

\section{Relaxation of the Magnetic Field\label{sec:mag_relax}}

\begin{figure}
\begin{center}
\plotone{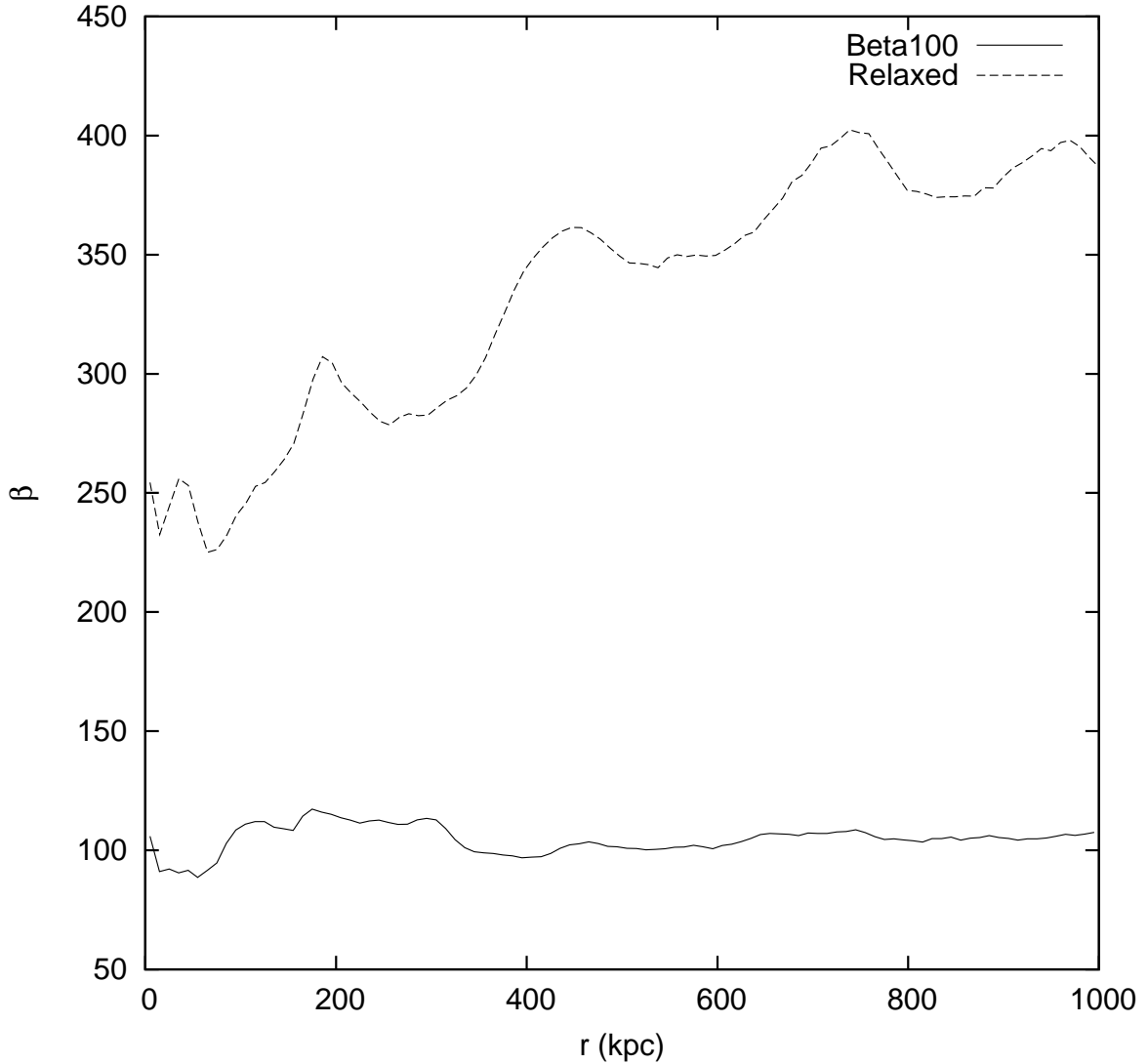}
\caption{Radial profiles of the plasma $\beta$ for the beginning of the relaxed simulation and the {\it Beta100} simulation.\label{fig:relaxed_beta}}
\end{center}
\end{figure}

\begin{figure}
\begin{center}
\plotone{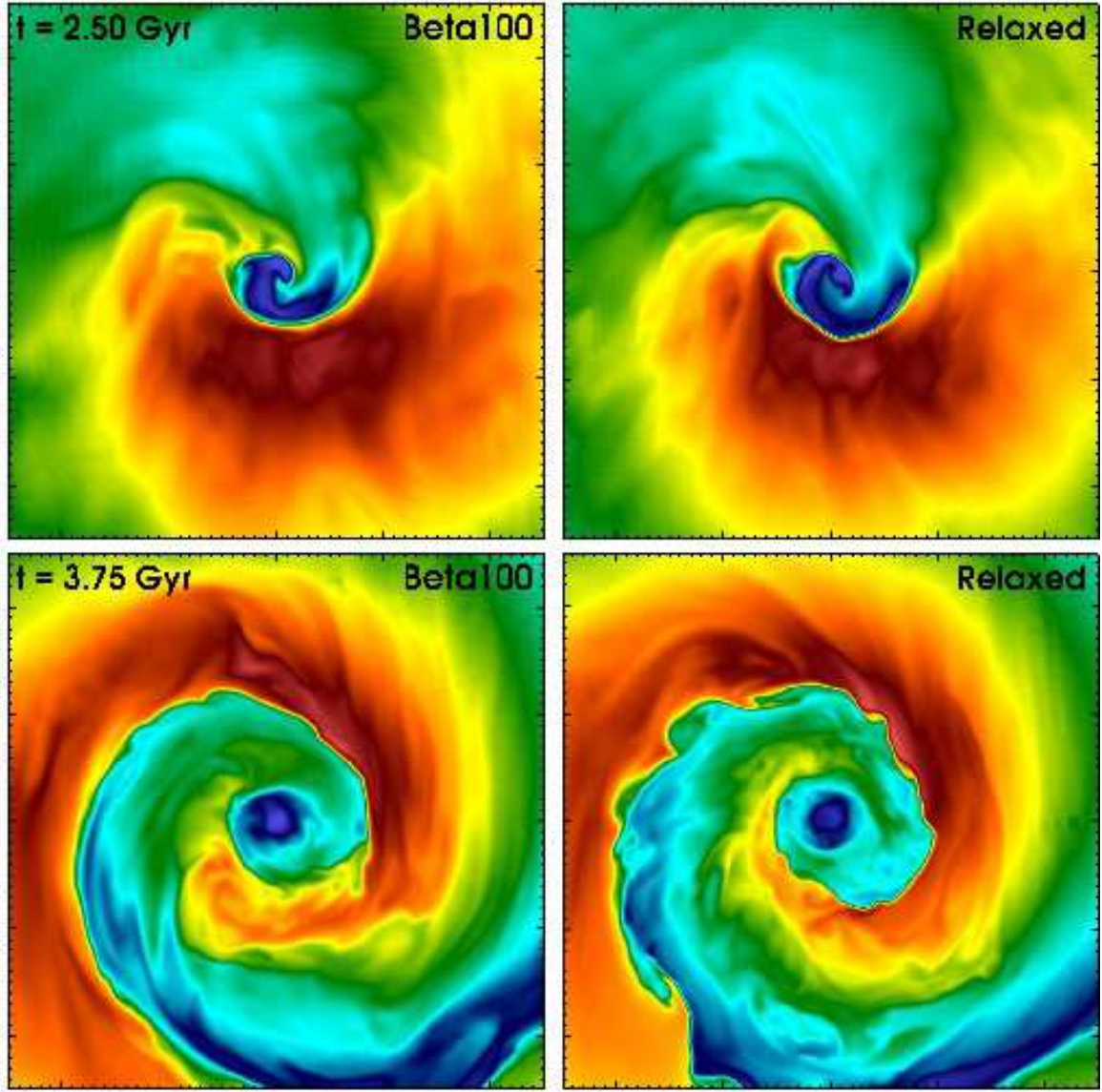}
\caption{Slices through the gas temperature in keV for the {\it Beta100} and relaxed simulations at the epochs $t$ = 2.5~Gyr (top panels) and $t$ = 3.75~Gyr (bottom panels). Each panel is 500 kpc on a side. Major tick marks indicate 100~kpc distances.\label{fig:relaxed_temp}}
\end{center}
\end{figure}

\begin{figure}
\begin{center}
\plotone{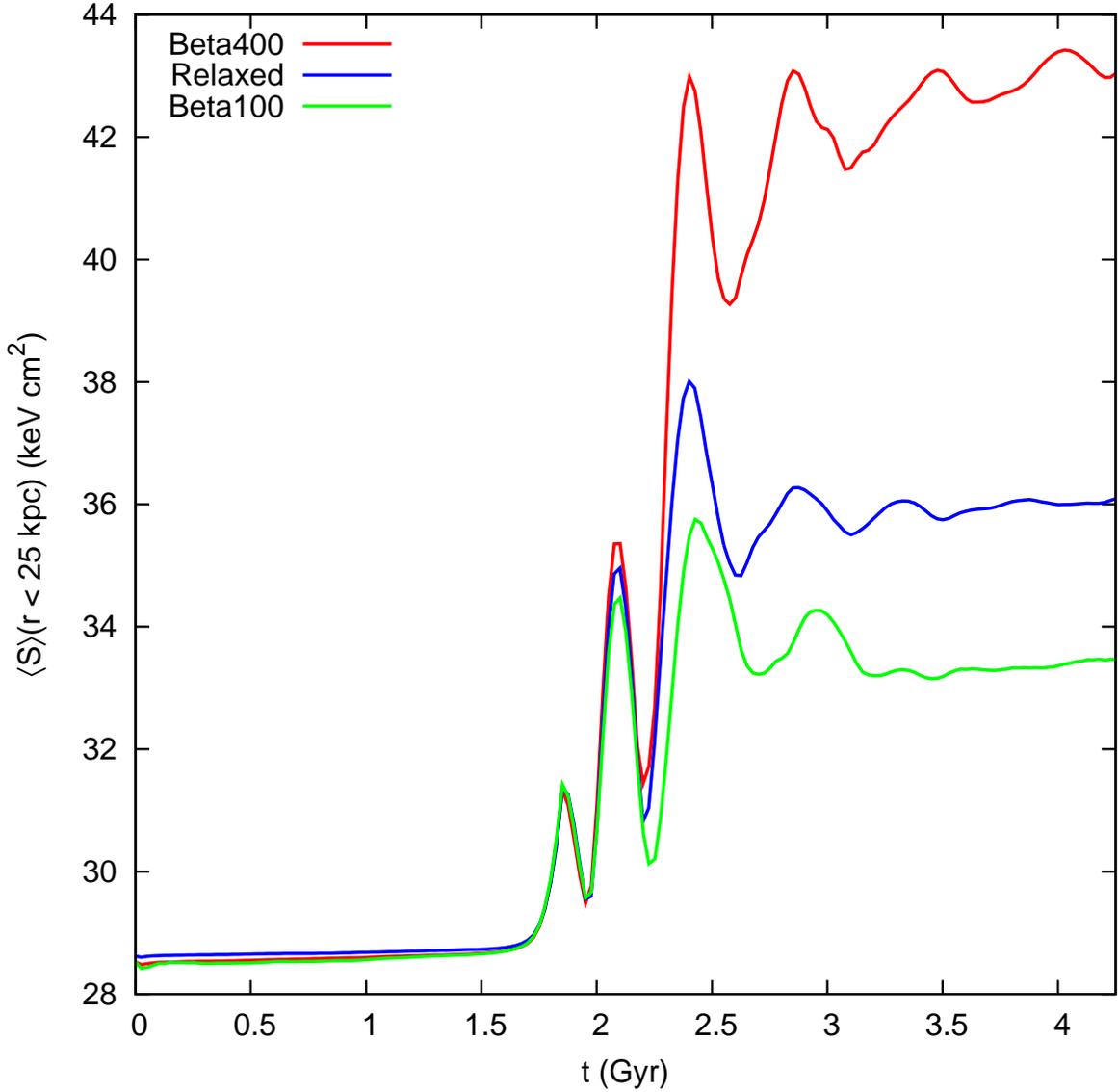}
\caption{Evolution of the average entropy within a radius of 25~kpc for the simulation with a relaxed magnetic field with initial $\beta$ = 100, compared to the {\it Beta100} and {\it Beta400} simulations.\label{fig:relaxed_entr}}
\end{center}
\end{figure}

\begin{figure*}
\begin{center}
\plotone{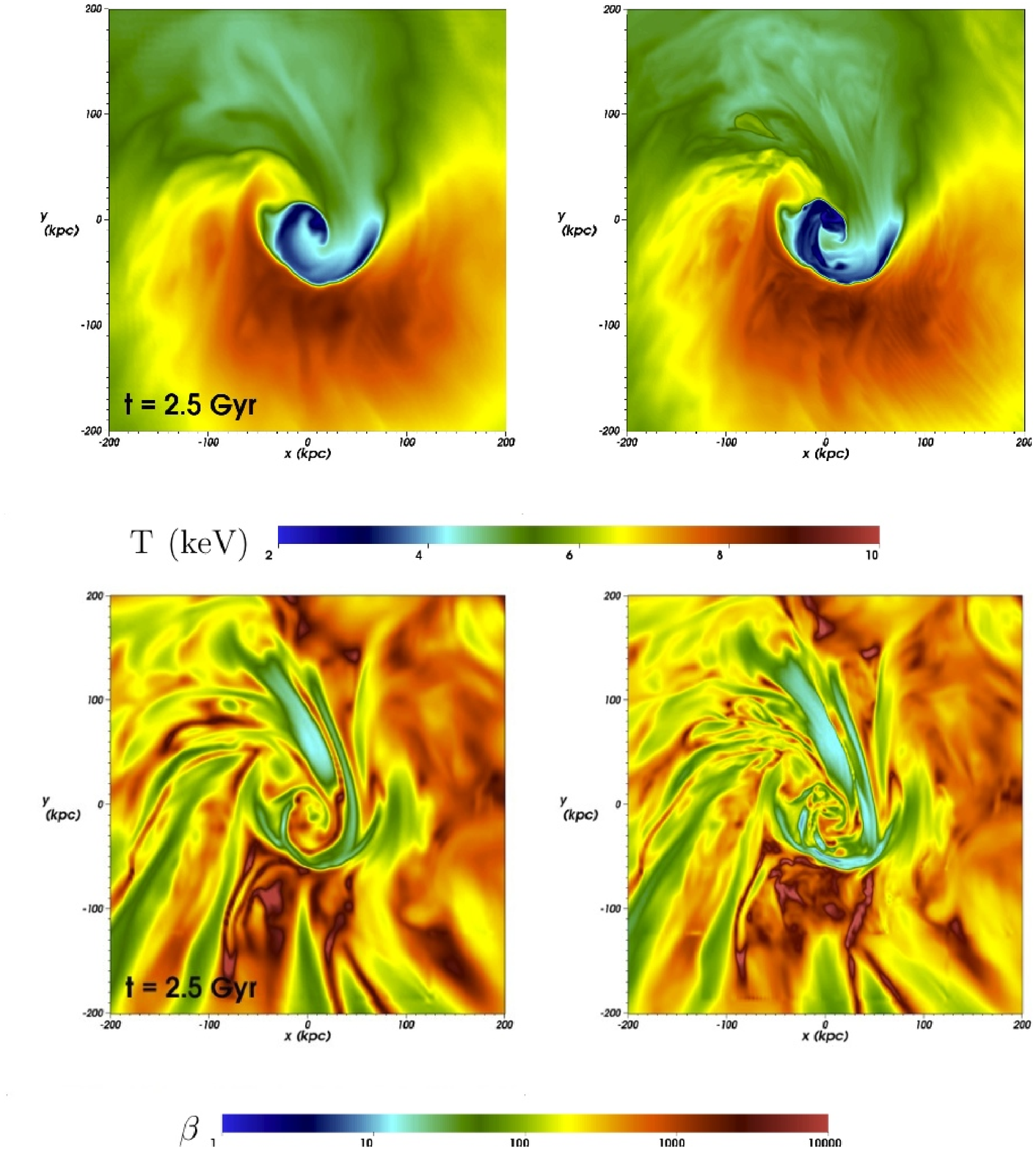}
\caption{Slices through the temperature and plasma beta for simulations of differing resolution, at the epoch $t$ = 2.5~Gyr. The simulation began with $\beta = 100$ and was relaxed for 6~Gyr before the encounter with the subcluster. Left: $\Delta{x} = 2$~kpc. Right: $\Delta{x} = 1$~kpc.\label{fig:res_test_t2.5}}
\end{center}
\end{figure*}

\begin{figure*}
\begin{center}
\plotone{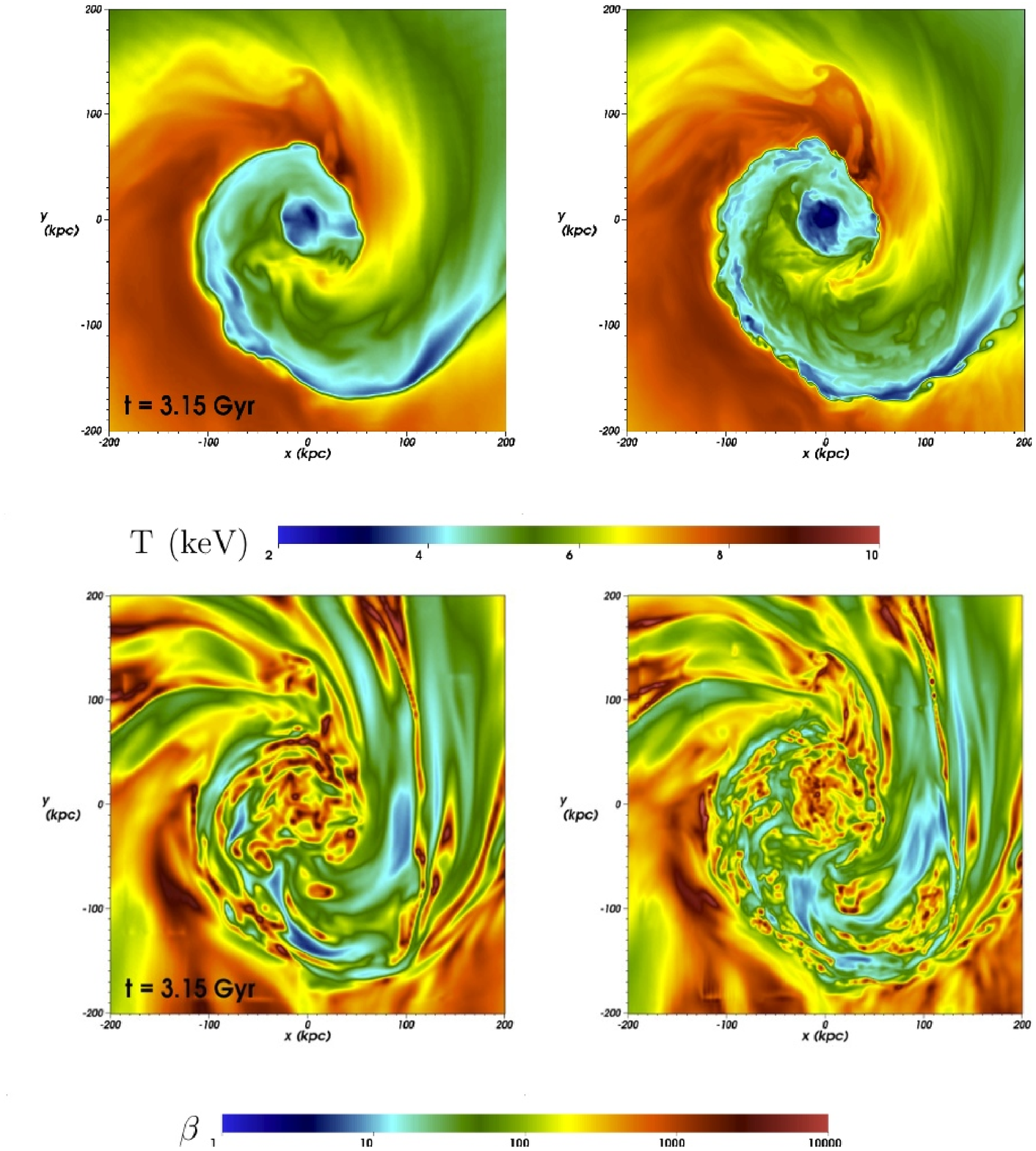}
\caption{Slices through the temperature and plasma beta for simulations of differing resolution, at the epoch $t$ = 3.15~Gyr. The simulation began with $\beta = 100$ and was relaxed for 6~Gyr before the encounter with the subcluster. Left column: $\Delta{x} = 2$~kpc. Right column: $\Delta{x} = 1$~kpc.\label{fig:res_test_t3.15}}
\end{center}
\end{figure*}

\begin{figure*}
\begin{center}
\plotone{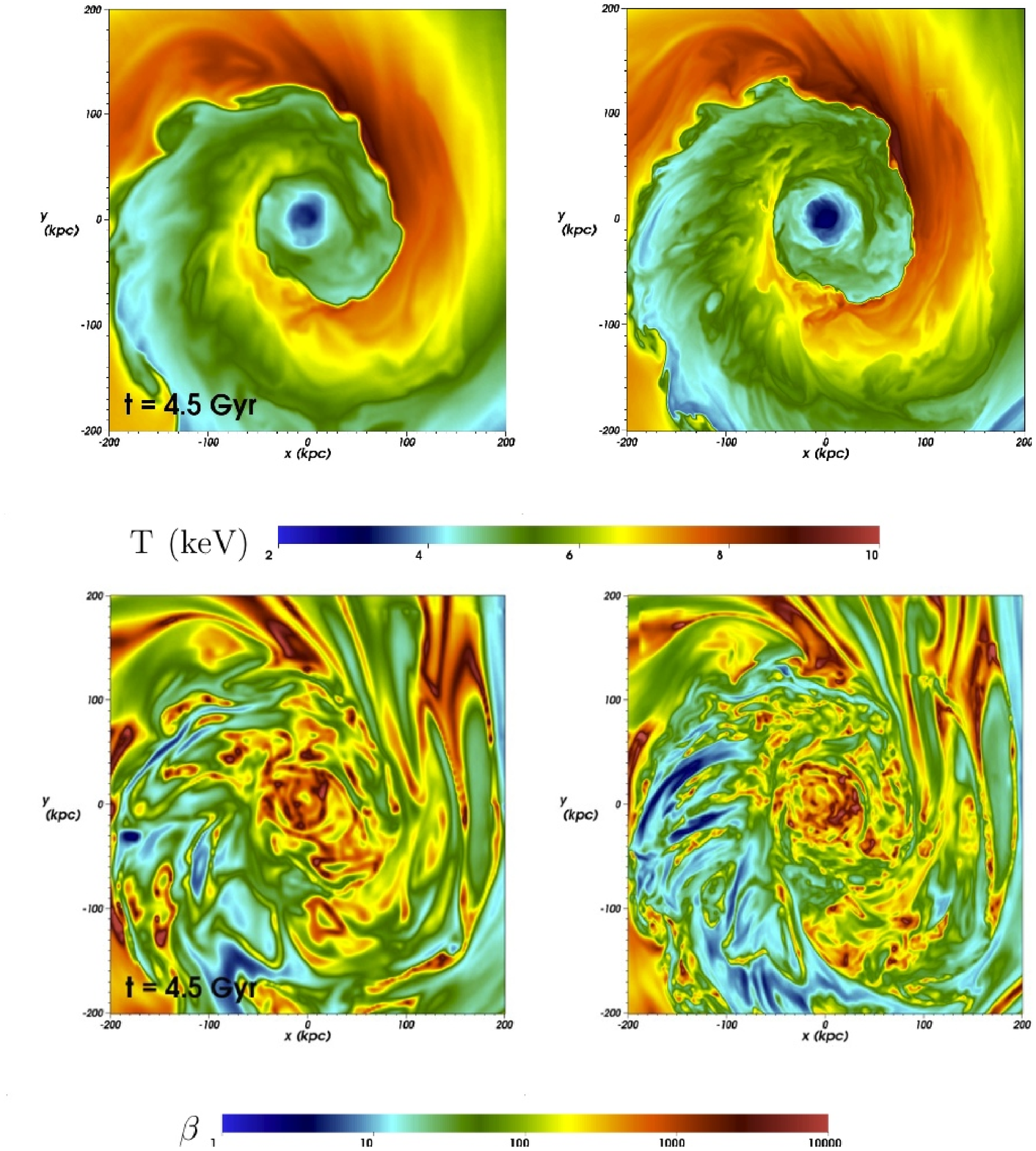}
\caption{Slices through the temperature and plasma beta for simulations of differing resolution, at the epoch $t$ = 4.5~Gyr. The simulation began with $\beta = 100$ and was relaxed for 6~Gyr before the encounter with the subcluster. Left column: $\Delta{x} = 2$~kpc. Right column: $\Delta{x} = 1$~kpc.\label{fig:res_test_t4.5}}
\end{center}
\end{figure*}

The initial magnetic fields in our simulations are not in an equilibrium configuration. On the one hand, the fields themselves are too dynamically weak to affect significantly the density and temperature structure of the gas, so we can assume that this structure will not change significantly until sloshing begins. However, if left to itself, the magnetic field of the cluster will rearrange itself to an equilibrium configuration, which may be a different configuration from which the simulation was started. In our default simulation setup, this process of relaxation is occuring while the subcluster is approaching the main cluster core and at the very beginning of the sloshing period. Since this period is not long, the relaxation is incomplete at the time of the passage of the subcluster, when the effects brought on by the encounter become much more important. To investigate the effects of this relaxation of the field on our results, we have run a second simulation with the same parameters as {\it Beta100}, except that the cluster has been allowed to evolve in isolation for 6~Gyr, when the magnetic field has approached an equilibrium configuration. Since this is the simulation with the strongest magnetic fields, we expect that the differences between the relaxed version and the original version will be most pronounced. We then begin the evolution of the simulation in the same fashion as the {\it Beta100} simulation. 

Figure \ref{fig:relaxed_beta} shows the radial profile of the plasma $\beta$ at the end of the relaxation of the single-cluster, compared with the radial profile of the plasma $\beta$ before any evolution. The final profile runs from $\beta \sim$~250 in the center of the cluster to $\sim$400 in the outskirts of the cluster. We might expect, based on our results from the simulations with varying initial $\beta$, that the effects caused by the evolution of this initial field due to sloshing would be intermediate between the {\it Beta100} and {\it Beta400} simulations. Figure \ref{fig:relaxed_temp} shows a comparison between temperature slices of the relaxed simulation and the {\it Beta100} simulation. In the relaxed-field simulation, the Kelvin-Helmholtz instabilities are still suppressed, but not as strongly as they are in the unrelaxed simulation. Indeed, the $t$ = 3.75~Gyr snapshot of the relaxed simulation in Figure \ref{fig:relaxed_temp} looks very similar to the one for the {\it Beta400} simulation shown in Figure \ref{fig:t3.75_temp}. Finally, Figure \ref{fig:relaxed_entr} shows a comparison between the evolution of the average entropy of the cool core within a radius of 25~kpc vs. time between the relaxed-field and {\it Beta100} simulations. We have also included the same evolution from the {\it Beta400} simulation for comparison. The increase in the entropy of the core due to mixing in the relaxed simulation is intermediate between the increase of the {\it Beta100} and {\it Beta400} simulations. 

We conclude from the results of this test that the primary differences between our {\it Beta100} simulation and its ``relaxed'' version are that the reduced magnetic field strengths present in the latter inhibit the growth of perturbations on the fronts less than in the former, and that mixing and the increase of entropy in the core are similarly less inhibited. This is in line with the conclusions from our simulations with varying initial $\beta$. 

\section{Resolution Test\label{sec:res_test}}

To test the robustness of our conclusions against the effects of varying resolution, we have performed a simulation with the same relaxed-field, $\beta = 100$ setup described in the previous section, but with a finest cell size of $\Delta{x} \sim 1$~kpc, half the cell size of the set of simulations described in this work. Increasing the resolution lowers the effective numerical viscosity of the simulation, allowing for smaller perturbations to be resolved and grow along the front surfaces. It also resolves smaller-scale fluctuations in the magnetic field, making it more difficult to maintain a coherent tangential field structure across the front surfaces. Both effects may affect our conclusions. 

Figures \ref{fig:res_test_t2.5} through \ref{fig:res_test_t4.5} show slices through the temperature (in keV) and the plasma $\beta$ parameter for the epochs $t$ = 2.5, 3.15, 3.75, and 4.5~Gyr after the beginning of the simulation for the relaxed-field initial setup with resolutions of $\Delta{x} \sim 2$~kpc and $\Delta{x} \sim 1$~kpc. The overall temperature structure, as well as the degree of field amplification, is very similar between the two simulations. However, two differences between the simulations are immediately apparent. The higher resolution-simulation allows for smaller fluctuations that ``wrinkle'' the front surface, and in some cases appear as small-scale K-H ripples (see Figure \ref{fig:res_test_t3.15} for a clear example). The result is that a front that was very smooth in the lower-resolution simulation is not smooth in the higher-resolution case. The second difference between the two simulations is that the magnetic field is much more tangled in the higher-resolution case. 

Regarding the inability of the fields to suppress K-H instabilities in the higher-resolution simulation, we note that the places where we see the appearance of ripples in the fronts when we go to higher resolution are the places where even in the lower-resolution simulation we note that the magnetic layers are not particularly strong or perfectly aligned with the front. This inhibits their effectivness in suppressing these instabilities. In these regions the suppression (or lack thereof) of instabilties along the fronts appears to be a resolution effect. In all the lower-resolution simulations, these were the cold fronts that were most susceptible to the growth of perturbations, due to the high shear velocities along these fronts and the weaker magnetic field. The lower-radii fronts at both resolutions are much smoother, due to the stronger magnetic field strengths and the lower shear velocities across these fronts. These dense, bright fronts are most prominent in X-ray observations of clusters, and for them our conclusions about the stability imparted to the front surfaces due to magnetic fields should still be valid. However, if the ICM has a significant viscosity, its effect will be to suppress instabilities and preserve the smoothness of the cold fronts, which may make this comparison academic. The effect of viscosity on cold fronts in a magnetized ICM will be the subject of a future paper.

\end{document}